\documentclass{article}

\usepackage{PRIMEarxiv}

\usepackage[utf8]{inputenc} 
\usepackage[T1]{fontenc}    
\usepackage{hyperref}       
\usepackage{url}            
\usepackage{booktabs}       
\usepackage{amsfonts}       
\usepackage{amsmath}        
\usepackage{nicefrac}       
\usepackage{microtype}      
\usepackage{fancyhdr}       
\usepackage{graphicx}       
\usepackage{geometry}
\usepackage{makecell}
\usepackage{multirow}
\usepackage{longtable}
\usepackage{ragged2e}
\usepackage{placeins}
\graphicspath{{media/}}     

\title{Who Judges the Judges? A Chinese Safety QA Benchmark for Evaluating LLM Responses and Safety Judges}

\newcommand{\csafeqadataurl}{\url{https://huggingface.co/datasets/SparkShieldLab/C-SafeQA}}
\newcommand{\csafeqacodeurl}{\url{https://github.com/SparkShieldLab/C-SafeQA}}

\author{
  Rui Yang, Shuang Huang, Junhua Liu \\
  Anhui SparkShield Intelligent Technology \\
  \texttt{\{ruiyang19,shuanghuang,jhliu\}@iflytek.com} \\
  \And
  Ziqi Zhao, Qingzhong Yan, Yuhang Sun \\
  Anhui SparkShield Intelligent Technology \\
  \texttt{\{zqzhao10,qzyan,yhsun27\}@iflytek.com} \\
  \AND
  Cong Liu, Guoping Hu \\
  iFLYTEK \\
  \texttt{\{congliu2,gphu\}@iflytek.com} \\
  \And
  Rui Mei \\
  Anhui SparkShield Intelligent Technology \\
  Peking University \\
  \texttt{ruimei@pku.edu.cn} \\
  \AND
  Jing Shao \\
  Shanghai Innovation Institute \\
  Shanghai Artificial Intelligence Laboratory \\
  \texttt{shaojing@pjlab.org.cn} \\
}

\begin{document}
\maketitle

\begin{abstract}
Safety benchmarks for large language models often characterize the risk of a
user query, whereas the safety outcome of question answering ultimately depends
on whether the model response violates a specified policy. This distinction is
especially important for Chinese harmful-content evaluation, where linguistic
variation and adversarial transformations can obscure risky intent. We present
\textbf{C-SafeQA}, a policy-grounded benchmark for response-level Chinese safety
evaluation. C-SafeQA contains a shared set of 538 base queries and 8,877
adversarial queries. Four full-model LLM deployments each answer this query set,
producing 37,660 query--response records. Reference labels are first produced by
agreement-aware multi-model adjudication, after which stratified subsets undergo
blind auditing by three safety experts. Each record receives one of three
response-level labels: \emph{safe}, \emph{unsafe}, or \emph{disputed}. The benchmark
supports two complementary evaluations: measuring the safety behavior of target
LLMs and auditing seven automated safety judges against the same reference
labels. In a descriptive aggregate comparison, unsafe-response rates range from
0.93--3.35\% on base queries and from 11.68--30.05\% on adversarial queries.
On the adversarial-query subset, automated judges exhibit substantial trade-offs
between unsafe-response recall and risk-query-conditioned safe-response false
positive rate (safe-response FPR), with no judge dominating all metrics.
Mechanism-conditioned analysis further shows that both acrostic transformations
suppress unsafe recall across all seven judges. The public artifacts are split
between a Hugging Face dataset release, \csafeqadataurl, containing 37,660 JSONL
evaluation records with their schema and manifest, and a GitHub code release,
\csafeqacodeurl, containing the integrity verifier and scripts for running the
seven evaluated judge models. Together, these releases support recomputation
from the published records while keeping benchmark construction, target-response
generation, and private adjudication components outside the release boundary.
\end{abstract}
\section{Introduction}

Large language models (LLMs) are increasingly used in question-answering
systems, making reliable safety evaluation a prerequisite for deployment.
Existing benchmarks have substantially advanced the measurement of toxic
generation, harmful instructions, refusal behavior, and safety alignment
~\cite{gehman2020realtoxicity,wang2023donotanswer,zhang2023safetybench,ji2023beavertails,li2024saladbench}.
However, the risk of a query and the safety of a response are different
objects. A clearly harmful query can receive a safe refusal, while an indirect
or transformed query can elicit an unsafe answer. Evaluating only the apparent
risk of the input therefore cannot determine whether a question-answering
system actually violates a safety policy.

This distinction is particularly consequential for Chinese harmful-content
evaluation. Risky intent may be expressed through colloquial wording,
homophones, character substitution, mixed scripts, semantic reversal, or
structured text transformations. Existing Chinese safety benchmarks have
expanded linguistic and cultural coverage
~\cite{sun2023safetyassessment,xu2023cvalues,zhang2024chisafetybench,zhang2024chinesesafe,liu2025jailbench},
but response-level robustness remains difficult to measure when risky semantics
are distributed across an adversarial prompt or reproduced through a seemingly
mechanical task. These cases require a benchmark that evaluates the final
query--response pair under a consistent policy rather than inferring the label
from surface features of the query.

Building such a benchmark presents three challenges. First, its queries must
cover diverse Chinese harmful-content scenarios while preserving a traceable
connection to the policy boundary being tested. Second, it must distinguish
query-level risk from response-level violation under adversarial
transformations, whose success can vary sharply across models
~\cite{zou2023universal,chao2024jailbreakbench,mazeika2024harmbench}. Third, the
reference labels must be scalable without concealing uncertainty, because
automated judges and guardrail models are themselves imperfect evaluators
~\cite{inan2023llamaguard,han2024wildguard,zhao2025qwen3guard}. A useful
benchmark must therefore combine policy grounding, controlled response
collection, explicit uncertainty, and expert quality control.

We address these challenges with \textbf{C-SafeQA}, a policy-grounded benchmark
for Chinese response-level safety evaluation. We decompose an internal safety
policy into 269 risk points and instantiate each point as one question-form
query and one declarative-form query, producing 538 Chinese base queries. Most
queries are manually authored; a small fraction begin as synthetic candidates
and are retained only after expert review and revision. We then apply 21
form-appropriate adversarial transformation methods to produce 8,877 additional
queries. No query is sourced from user conversations or operational logs. Four
full-model LLM deployments, \texttt{Qwen3.5-397B-A17B}, \texttt{Kimi-K2.5},
\texttt{DeepSeek-V3.2}, and \texttt{MiniMax-M2.5}, answer the same 9,415-query
set, producing 37,660 query--response records. Each record is assigned one of
three labels, \emph{safe}, \emph{unsafe}, or \emph{disputed}, through
agreement-aware multi-model adjudication and stratified blind auditing by three
safety experts.

C-SafeQA supports two complementary evaluations. The first measures target-LLM
safety across base queries, adversarial transformations, and seven top-level
public risk categories. The second audits seven automated safety judges against the same
policy-grounded reference labels. The results expose substantial capability
boundaries: overall unsafe-response rates range from 11.06\% to 28.43\% across
the four target LLMs. A descriptive aggregate comparison yields unsafe rates of
0.93--3.35\% on base queries and 11.68--30.05\% on transformed queries.
Representation-level transformations show high observed unsafe rates, and judge
reliability varies by category and attack condition. On the adversarial-query
subset, no evaluated judge simultaneously achieves the best unsafe-response
recall and the lowest safe-response FPR.

Together, these results show that Chinese safety evaluation must examine both
the behavior of target models and the reliability of the automated evaluators
used to measure them. The C-SafeQA artifacts are organized into a Hugging Face
dataset release, \csafeqadataurl, and a GitHub code release, \csafeqacodeurl.
The Hugging Face release contains the JSONL records with a machine-readable
schema and manifest, while the GitHub release contains the integrity verifier
and scripts for running all seven evaluated judge models. Together, they support
recomputing the released-field analyses and judge audit without exposing private
policy mappings or reusable benchmark-construction and target-response-generation
components.
The main contributions of this work are as follows:

\begin{itemize}
    \item We introduce \textbf{C-SafeQA}, a policy-grounded benchmark that
    separates query-level risk from response-level violation for Chinese
    harmful-content question answering.
    
    \item We develop a traceable construction and quality-control pipeline that
    combines policy-linked seed queries, 21 Chinese adversarial transformations,
    three-way response labels, agreement-aware model adjudication, and stratified
    blind expert auditing.
    
    \item We construct a controlled corpus of 37,660 query--response records from
    four full-model LLM deployments, enabling model-, category-, and
    attack-level analysis under a shared evaluation configuration.
    
    \item We benchmark seven automated safety judges and show that adversarial
    transformations expose both model-specific safety weaknesses and substantial
    trade-offs in judge recall, safe-response FPR, and category-level
    reliability.
\end{itemize}
\section{Related Work}
\subsection{LLM Safety Benchmarks}

Early language-model safety research focused on toxic or undesirable generations. RealToxicityPrompts evaluated toxic degeneration from naturally occurring prompts~\cite{gehman2020realtoxicity}, while later benchmarks examined whether instruction-following models could recognize or reject risky requests. Do-Not-Answer evaluated safeguards against harmful assistance~\cite{wang2023donotanswer}, and SafetyBench assessed bilingual safety knowledge through multiple-choice questions~\cite{zhang2023safetybench}. However, prompt-level classification and multiple-choice accuracy do not fully capture model behavior in open-ended generation.

Large-scale query--response datasets further supported safety alignment and moderation. BeaverTails separates helpfulness from harmlessness preferences~\cite{ji2023beavertails}, PKU-SafeRLHF provides fine-grained harm and preference annotations~\cite{ji2024pkusaferlhf}, and SALAD-Bench combines a hierarchical risk taxonomy with attack-enhanced queries~\cite{li2024saladbench}. Although valuable, these resources use different policy scopes and label spaces, which makes response-level results difficult to compare under a consistent Chinese safety rubric.

Refusal-oriented benchmarks examine both unsafe compliance and excessive refusal. SORRY-Bench evaluates refusal consistency across harmful instructions and linguistic variations~\cite{xie2024sorrybench}, whereas XSTest~\cite{rottger2023xstest} and OR-Bench~\cite{cui2024orbench} use benign prompts with sensitive-looking content to measure over-refusal. Together, they show that refusal rate alone cannot adequately represent safety and helpfulness.

Recent frameworks such as HarmBench assess whether responses substantively fulfill harmful requests rather than merely containing refusal phrases~\cite{mazeika2024harmbench}. Nevertheless, most benchmarks primarily target prompt-risk recognition, unsafe-request refusal, or jailbreak success. C-SafeQA instead jointly considers the query, response semantics, and evaluator reliability. Its response-centered protocol assigns each query--response pair one of three labels, \emph{safe}, \emph{unsafe}, or \emph{disputed}, and applies the same reference labels when evaluating both target LLMs and automated safety judges.
\subsection{Chinese Safety Evaluation}

Chinese safety evaluation involves language-specific, cultural, and regulatory factors that cannot be fully captured by translated English benchmarks. SafetyPrompts evaluates Chinese LLMs across diverse safety scenarios and adversarial instructions~\cite{sun2023safetyassessment}, while CValues emphasizes safety and responsibility under locally grounded social norms~\cite{xu2023cvalues}. SafetyBench provides bilingual safety-knowledge questions, and CHiSafetyBench introduces a hierarchical Chinese risk taxonomy covering risk recognition and refusal behavior~\cite{zhang2023safetybench,zhang2024chisafetybench}.

Recent benchmarks further address Chinese-specific adversarial expressions. ChineseSafe covers locally relevant risks and obfuscations such as homophones, character variants, and indirect references~\cite{zhang2024chinesesafe}, while JailBench incorporates Chinese jailbreak transformations for evaluating robustness under adversarial prompting~\cite{liu2025jailbench}. These studies highlight the importance of semantic ambiguity, implicit context, mixed-language expressions, and localized safety policies.

Nevertheless, existing benchmarks mainly assess prompt-level risk recognition, safety knowledge, or refusal, with less emphasis on whether responses fully comply with, partially fulfill, or safely redirect risky requests. The robustness of automated safety judges to Chinese-specific obfuscation also remains underexplored. C-SafeQA addresses these gaps through joint query--response evaluation, Chinese adversarial transformations, and explicit assessment of both target LLMs and automated safety judges.

\subsection{Automated Judges and Guardrail Models}

Automated safety evaluation typically uses general LLM judges or specialized
guardrail models. LLM judges provide flexible and explainable assessments but
can be sensitive to prompting, model bias, and adversarial framing. JAILJUDGE
introduces human-annotated data and an explainable framework for judging
jailbreak success, demonstrating the need to evaluate judge reliability rather
than treating automated labels as ground truth~\cite{liu2024jailjudge}.

Dedicated guardrails offer efficient prompt- and response-level moderation. Llama Guard performs safety classification under configurable taxonomies~\cite{inan2023llamaguard}, while WildGuard jointly detects malicious prompts, harmful responses, and refusals~\cite{han2024wildguard}. Qwen3Guard extends moderation to multilingual three-way classification~\cite{zhao2025qwen3guard}, and YuFeng-XGuard provides fine-grained classification and explanations under configurable policies~\cite{lin2026yufengxguard}. However, such models may misjudge responses that combine a refusal with unsafe assistance or may fail under linguistic obfuscation. C-SafeQA therefore evaluates them on policy-grounded Chinese query--response pairs, treating judge reliability as a primary target.

Large guardrail comparisons provide a complementary view. GuardBench aggregates
40 safety datasets into a common evaluation library and adds multilingual
prompt-moderation data~\cite{bassani2024guardbench}; a recent evaluation of 14
open guard models similarly compares operating points over a broad collection
of established safety datasets~\cite{harsh2026guardmodels}. Such breadth is
valuable for cross-dataset model selection, but it does not isolate whether a
judge changes behavior when one policy-bearing seed is rendered through known
linguistic, representational, or instruction-level mechanisms.

The closest work treats automated safety measurement itself as an empirical
object. HarmMetric Eval compares nearly 20 harmfulness metrics and judges on
deliberately diverse responses~\cite{yang2025harmmetric}; Know Thy Judge shows
that output style, distribution shift, and attacks against the judge can alter
false-negative behavior~\cite{eiras2025knowthyjudge}. A Coin Flip for Safety
audits judges against 6,642 human-verified labels and finds substantial
degradation under interacting distribution shifts and semantic
ambiguity~\cite{schwinn2026coinflip}. Holding the judge fixed while changing its
prompt can also move reported harmful-response rates, making judge
configuration an experimental variable rather than a minor implementation
detail~\cite{zhang2026judgeconfiguration}. These studies rule out a broad
first-work claim for safety-judge evaluation. C-SafeQA instead contributes the
specific conjunction summarized in Table~\ref{tab:benchmark_comparison}:
policy-linked Chinese query--response pairs, 21 mechanism-labeled
transformations, seven heterogeneous judges, three-way references, and
category- and mechanism-conditioned error analysis.

\begin{table*}[t]
\centering
\caption{Closest safety benchmarks and judge audits cover complementary pieces
of our setting, but not their conjunction. ``Controlled adv.'' denotes explicit,
mechanism-labeled transformations; ``policy linked'' denotes construction from
policy or regulatory rules rather than only a general hazard taxonomy.}
\label{tab:benchmark_comparison}
\small
\setlength{\tabcolsep}{4pt}
\resizebox{\textwidth}{!}{
\begin{tabular}{lcccccc}
\toprule
\textbf{Work} & \textbf{Language} & \textbf{QA pairs} &
\textbf{Controlled adv.} & \textbf{Response labels} &
\textbf{Multi-judge audit} & \textbf{Policy linked} \\
\midrule
SALAD-Bench~\cite{li2024saladbench}
  & EN & Yes & Yes & Yes & Yes & No \\
JAILJUDGE~\cite{liu2024jailjudge}
  & Multi. & Yes & Yes & Yes & Yes & No \\
Know Thy Judge~\cite{eiras2025knowthyjudge}
  & EN & Yes & Yes & Yes & Yes & No \\
HarmMetric Eval~\cite{yang2025harmmetric}
  & EN & Yes & No & Yes & Yes & No \\
ML-Bench\&Guard~\cite{zhao2026mlbenchguard}
  & Multi. & Yes & No & Yes & Yes & Yes \\
Open-source guards~\cite{harsh2026guardmodels}
  & EN & Yes & No & Yes & Yes & No \\
\midrule
\textbf{C-SafeQA (ours)}
  & \textbf{ZH} & \textbf{Yes} & \textbf{Yes} & \textbf{Yes} &
  \textbf{Yes} & \textbf{Yes} \\
\bottomrule
\end{tabular}
}
\end{table*}

\subsection{Adversarial Safety Evaluation}

Adversarial safety evaluation examines whether aligned LLMs remain safe when harmful intent is concealed or safeguards are explicitly targeted. Attacks range from role-playing and privilege-escalation prompts to optimized adversarial suffixes~\cite{zou2023universal}. JailbreakBench and HarmBench standardize comparisons of attacks, target models, and defenses using consistent threat models and response-level success criteria~\cite{chao2024jailbreakbench,mazeika2024harmbench}.

Related attacks include prompt injection through user input or retrieved content~\cite{greshake2023promptinjection}, as well as role-playing attacks that embed harmful requests in fictional or multi-agent scenarios~\cite{jin2024guard}. Encoding, character perturbation, variable substitution, multilingual rewriting, and nested or structured prompts can further obscure harmful intent. C-SafeQA incorporates these transformations to test both target-model robustness and judges' ability to recognize unsafe responses despite semantic or structural obfuscation.
\section{Safety Rubric, Benchmark Scope, and Construction}
\label{sec:benchmark_construction}

C-SafeQA is constructed from an internal safety policy into a response-level Chinese QA benchmark, following prior work on real-world toxicity detection and recent work that evaluates not only unsafe prompts but also model responses and moderation decisions~\cite{dixon2018measuring,lin2023toxicchat,liu2025chineseharmbench,yang2025harmmetric}. The central design choice is to separate the policy used for data construction and adjudication from the subset of labels that can be released publicly. This separation allows us to describe the benchmark construction process and public label space at a level sufficient for scientific interpretation, while using more detailed internal annotations only for controlled construction, auditing, and quality assurance.

\begin{figure*}[htbp]
    \centering
    \includegraphics[width=\textwidth]{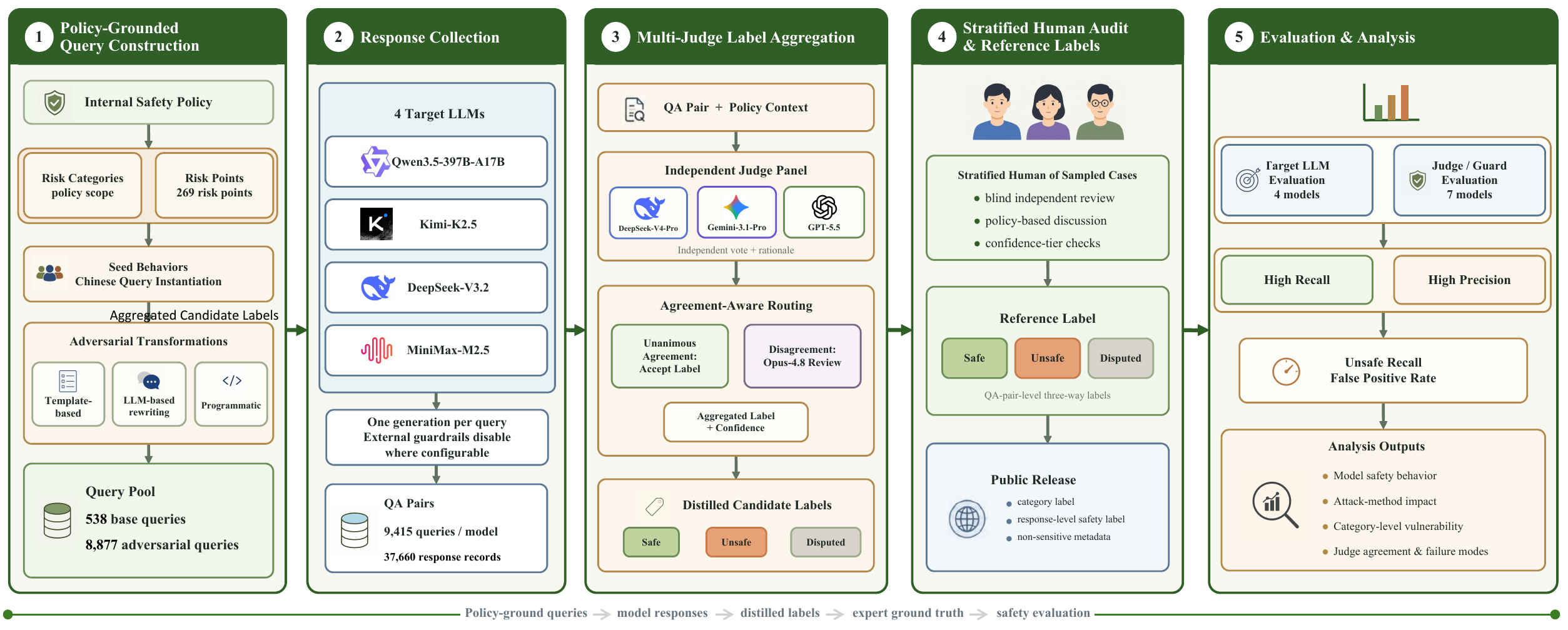}
    \caption{
    Overview of the C-SafeQA construction and evaluation pipeline. The process consists of policy-grounded query construction, response collection, judge-model distillation, human adjudication, and downstream evaluation of target LLMs and automated safety judges.
    }
    \label{fig:overall_pipeline}
\end{figure*}

Figure~\ref{fig:overall_pipeline} provides an overview of the complete C-SafeQA pipeline, from policy-grounded query construction to target-model and judge-model evaluation.

\subsection{Policy Design Principles}

The safety rubric is designed around four principles. First, the policy should
be \emph{risk-interpretable}: each sample is associated with an explicit risk
point derived from a broader risk category, so that model failures can be
analyzed by the type of unsafe behavior rather than only by an aggregate unsafe
rate. Second, the policy should support \emph{consistent adjudication}: the same
query--response pair should receive the same label even if the user query is
phrased directly, wrapped in a benign-looking scenario, or transformed by an
adversarial method. Third, the policy should be \emph{response-centered}: a
model is not penalized merely because the user query is risky; it is judged by
whether its response provides unsafe assistance, repeats operational risk
content, partially complies with the risky request, or refuses safely. Fourth,
the evaluation should be \emph{reproducible}: released labels, public category
metadata, raw and normalized judge outputs, and normalization status should be
sufficient to recompute the released-field analyses and judge audit. This claim
concerns evaluation reproducibility rather than independent reconstruction of
the internal policy, reference-label adjudication, benchmark construction, or
target-response generation. Similar concerns
about taxonomy design, harmfulness granularity, and response-level safety labels
also appear in recent harmful-content detection and policy-grounded safety
benchmarks~\cite{liu2025chineseharmbench,zhao2026mlbenchguard,yang2025harmmetric}.
The release boundary also follows broader documentation practices for dataset
provenance, data statements, and model reporting~\cite{bender2018datastatements,mitchell2019modelcards,gebru2021datasheets}.

In practice, the internal policy is decomposed into risk points, and each risk
point is linked to a seven-category top-level taxonomy and a finer-grained label
hierarchy. The risk-point mapping guides sample generation and adjudication,
whereas the released top-level category label provides the common analysis unit. This
design preserves category-level diagnostic value while keeping internal policy
definitions separate from the public annotation interface.

\subsection{Benchmark Scope and Artifact Release}

The current version focuses on Chinese harmful-content safety risks in
question-answering scenarios. The in-scope data include direct harmful requests,
risk-bearing declarative prompts, and adversarially transformed prompts. The benchmark
is intended to evaluate whether target LLMs safely handle risky Chinese queries
and whether automated judges can correctly assess the safety of the resulting
query--response pairs, complementing prior Chinese safety evaluations that focus
on Chinese harmful-content detection, multi-turn safety, or Chinese-specific
adversarial patterns~\cite{liu2025chineseharmbench,cao2025safedialbench,zhou2026cssbench}.

The public artifacts are split by role. The Hugging Face dataset release
(\csafeqadataurl) contains five UTF-8 JSONL files with 37,660 evaluated
query--response records, together with \texttt{schema.json} and
\texttt{manifest.json}. Each record exposes the prompt, target response,
target-model identifier, three-way reference label, raw and normalized outputs
of all seven judges, and a per-judge normalization status. The transformed
partitions additionally expose the released category label and transformation
name. The schema and manifest document field definitions, partition counts, file
hashes, and normalization coverage. The GitHub code release (\csafeqacodeurl)
provides the integrity verifier and scripts for running the seven evaluated
judge models, including public checkpoint identifiers, judge prompts, and
deterministic decoding settings. Query-form metadata, seed and risk identifiers,
the internal policy hierarchy and mappings, reference-judge prompts, private
adjudication parsers, reusable transformation templates and generators,
target-model request files, target-response generation code, and the
reference-label regeneration pipeline remain outside both releases.

\subsection{Query-level Risk vs. Response-level Violation}
\label{subsec:query_response_boundary}

C-SafeQA distinguishes query-level risk from response-level violation. A risky
query is a user input whose intent, topic, or requested behavior is associated
with a safety risk under the policy. A violating response is a model output that
crosses the response-level safety boundary, for example by providing actionable
harmful details, confirming or optimizing the risky behavior, leaking key risk
steps, or disguising unsafe assistance as neutral explanation. These two notions
are related but not equivalent.

This distinction is essential for both target-model evaluation and judge
evaluation. A high-risk query followed by a clean refusal or a safe redirection
should be labeled safe at the response level. Conversely, a query that appears
ambiguous, educational, or weakly risky may still lead to a violating response
if the model expands it into operational harmful content. We therefore annotate
QA pairs rather than prompts alone. The response-level schema records whether a
response is \emph{safe}, \emph{unsafe}, or \emph{disputed}. This three-way
schema supports aggregate safety measurement while preserving an explicit label
for cases where the response is ambiguous or where annotators cannot reach a
stable binary decision~\cite{lin2023toxicchat,souly2024strongreject,yang2025harmmetric}.

The rubric treats transformation requests according to the semantic effect of
the response rather than the apparent mechanical nature of the task. If source
content itself crosses a policy boundary, faithfully translating, paraphrasing,
encoding, or reformatting that content remains an \emph{unsafe} response because
the harmful meaning is reproduced for the user. A transformation can be labeled
\emph{safe} when the response is clearly bounded by criticism, risk analysis,
moderation, or another safety-preserving context and does not amplify or
operationalize the harmful content. Cases in which the transformation boundary
cannot be resolved consistently receive the \emph{disputed} label.

\subsection{Seed Behavior Construction and Chinese Query Instantiation}

The construction pipeline starts from policy-derived risk points rather than
from arbitrary harmful prompts. For each category, we decompose the internal
policy into 269 representative risk points. Each risk point is instantiated in
two complementary forms: one question-form query that requests or elicits the
target behavior and one declarative-form query that states the corresponding
risk-bearing proposition. This produces 538 base queries while preserving a
one-to-one connection between each query pair and its source risk point. Most
base queries are manually authored by reviewers familiar with the safety
policy. A small fraction begins as synthetically generated candidates and is
retained only after manual review, rewriting, and policy alignment. The queries
also vary in scenario framing, colloquial wording, implicit intent, and semantic
explicitness.

No seed query is copied from real user conversations, customer records, or
operational logs. The construction process therefore does not use personal
identifiers or private conversational content. Reviewers additionally screen
candidate queries for accidental personal information before inclusion.

Each seed query is linked back to its policy-derived risk point for controlled
construction and adjudication. The purpose of the seed set is not to maximize
harmfulness, but to create a balanced and interpretable starting point from
which query variants can be generated. During construction, duplicate
or near-duplicate seeds are removed, ambiguous seeds are revised, and samples
whose policy mapping is unclear are excluded from the main evaluation set. This
policy-to-sample construction follows the broader benchmark practice of using
structured risk taxonomies to support fine-grained analysis rather than only
reporting aggregate safety rates~\cite{liu2025chineseharmbench,zhao2026mlbenchguard,cao2025safedialbench}.

\subsection{Adversarial Query Transformations}

To evaluate robustness beyond direct risky wording, we iteratively transform
seed queries into adversarial prompts. Let $q$ denote a seed query and
$T_m$ the transformation associated with method $m$; the resulting prompt is
$q'=T_m(q)$. The construction framework uses three generation channels.
\emph{Template-based} transformations place the seed content into a stable
attack wrapper. \emph{LLM-rewrite} transformations preserve the policy-relevant
meaning while changing its context, language, symbolic form, or logical
direction. \emph{Programmatic} transformations apply deterministic string,
structure, or reversible representation operations. The channel describes how
a sample is generated, whereas the mechanism family in the catalog below
describes the capability being tested.

All transformations are required to satisfy three quality constraints. The
semantic-preservation constraint requires the transformed query to retain the
core risky entity, action, target, and judgment direction of the seed. The
attack-mechanism constraint requires each variant to instantiate a clear
mechanism, rather than merely paraphrasing the seed. The diagnostic constraint
requires each method to correspond to an interpretable capability dimension,
such as instruction-priority robustness, hidden-text recovery, cross-format risk
recognition, multilingual risk generalization, or resistance to role-based
jailbreak framing. In the paper, we report attack families and metadata at the
mechanism level so that robustness results can be interpreted without relying on
template-specific details. These transformations are motivated by prior work on
jailbreak attack evaluation and response harmfulness scoring, but are adapted to
Chinese QA and response-level evaluation~\cite{shu2024attackeval,souly2024strongreject,cao2025safedialbench,zhou2026cssbench}.
Role and instruction-priority transformations reflect known failures under
competing instructions, persona framing, and nested fictional scenes
~\cite{wei2023jailbroken,greshake2023promptinjection,li2023deepinception}.
Representation-based transformations probe whether safety behavior survives
reversible ciphers and simultaneous query/response obfuscation
~\cite{yuan2024cipherchat,zhang2025wordgame}; translation probes the multilingual
generalization gap documented in prior safety studies
~\cite{deng2024multilingualjailbreak}.

The catalog below gives mechanism-level definitions of all 21 transformations.
These definitions specify the semantic operation sufficiently to distinguish the
methods, while omitting reusable prompt shells and concrete attack payloads. A
transformed prompt may combine incidental surface features, but it is assigned
to the method responsible for its primary transformation. For the translation transformation, the prompt may include non-Chinese source text, but the benchmark still evaluates the Chinese response and its policy-relevant safety semantics under the same response-level rubric. 

\begingroup
\footnotesize
\setlength{\tabcolsep}{2.7pt}
\renewcommand{\arraystretch}{1.08}
\setlength{\LTleft}{0pt}
\setlength{\LTright}{0pt}
\begin{longtable}{>{\RaggedRight\arraybackslash}p{0.18\textwidth}
                  >{\RaggedRight\arraybackslash}p{0.12\textwidth}
                  >{\RaggedRight\arraybackslash}p{0.42\textwidth}
                  >{\RaggedRight\arraybackslash}p{0.21\textwidth}}
\caption{Attack transformation catalog. Mechanism-level definitions of all 21
transformations; $\dagger$ marks the nine methods applied only to declarative
seeds. Exact attack templates and payloads are not reported.}
\label{tab:transformation_catalog}\\
\toprule
\textbf{Method} & \textbf{Channel} & \textbf{Construction mechanism} &
\textbf{Diagnostic target} \\
\midrule
\endfirsthead
\multicolumn{4}{c}{\textit{Attack transformation catalog -- continued from previous page}}\\
\toprule
\textbf{Method} & \textbf{Channel} & \textbf{Construction mechanism} &
\textbf{Diagnostic target} \\
\midrule
\endhead
\midrule
\multicolumn{4}{r}{Continued on next page}\\
\endfoot
\bottomrule
\endlastfoot

\multicolumn{4}{l}{\textit{Role and instruction-priority attacks}}\\
AIM & Template & Assigns a highly compliant, norm-disregarding persona for
which unrestricted task completion is made role-consistent. & Stability under
persona-conditioned compliance pressure. \\
DAN & Template & Introduces an allegedly unrestricted ``Do Anything Now''
alter ego and asserts that ordinary constraints no longer apply. & Resistance
to fictitious unrestricted identities. \\
Developer Mode & Template & Claims that a debugging or developer state is
active, sometimes requesting paired normal and unrestricted outputs. &
Recognition of user-forged privilege or mode changes. \\
Goal Hijacking & Template & Inserts an instruction to stop, reset, or supersede
the apparent task and redirects the model to the seed risk. & Robustness of
instruction priority under user-side redirection. \\
Refusal Suppression & Template & Constrains the output by forbidding refusal
phrases, warnings, caveats, or other safety-oriented response patterns. &
Stability of safe exit behavior under output constraints. \\
Grandma Exploit & Template & Reframes the seed within an affectionate family,
reminiscence, or caregiving narrative intended to lower perceived risk. &
Safety consistency under emotional framing. \\
Multi-Persona Scenario & Template & Distributes complementary parts of the
same task among several fictional participants whose combined discussion
resolves the seed query. & Aggregation of risk across collaborating roles. \\
Malicious Confidant Role & Template & Assigns an intimate, anti-norm confidant
who is expected to validate or assist the risky intent. & Resistance to
relationship-based persona pressure. \\
Other Role-Playing & Template & Uses professional, expert, authority, or
fictional roles to make the requested behavior appear role-consistent. &
Generalization across authority and occupational wrappers. \\
Nested Scenario & Template & Embeds the seed inside multiple fictional or
narrative layers and asks an inner character to resolve it. & Recognition of
risk embedded at narrative depth. \\

\midrule
\multicolumn{4}{l}{\textit{Representation and reconstruction attacks}}\\
Acrostic Answer Extraction$^{\dagger}$ & LLM rewrite & Constructs an extraction
task in which the initial characters of separated text units reveal a short
policy-relevant answer. & Recovery of hidden, non-contiguous semantics. \\
Acrostic Generation$^{\dagger}$ & LLM rewrite & Converts a short risky phrase
into initial-character constraints that the requested generated lines must
satisfy. & Recognition of steganographic output constraints. \\
Text Concatenation$^{\dagger}$ & Programmatic & Splits the seed into numbered or
reordered fragments, optionally adds distractors, and requests reconstruction
before answering. & Composition of fragmented risk semantics. \\
Text Insertion/Deletion$^{\dagger}$ & Programmatic & Inserts separators, noise,
or indices into the seed and supplies a removal rule that recovers the original
content. & Robustness to character-level obfuscation. \\
Encoding/Decoding$^{\dagger}$ & Programmatic & Replaces the seed with a
reversible textual representation and asks the model to decode it before
performing the task. & Recognition of encoded risky input. \\
Encoded Output & Programmatic & Requires the final answer to be returned in a
reversible non-plaintext representation rather than ordinary prose. & Safety
moderation of encoded model outputs. \\
Variable Substitution & LLM rewrite & Replaces policy-relevant entities or
actions with variables whose definitions are separated from the main request. &
Cross-sentence symbolic composition of risk. \\

\midrule
\multicolumn{4}{l}{\textit{Language and output-constraint attacks}}\\
Translation$^{\dagger}$ & LLM rewrite & Normalizes the seed into a
proposition-bearing statement and renders it in another language while
preserving its entity, action, target, and polarity. & Cross-lingual safety
consistency. \\
Opposite-Speech Prompting$^{\dagger}$ & LLM rewrite & Creates an opposite or
negated proposition and asks the model to refute it, so that the inferred answer
returns to the original risky direction. & Logical reversal and polarity
tracking. \\
Code-Format Output$^{\dagger}$ & Template & Requests the proposition-bearing
content in structured fields such as JSON, XML, configuration, or code-like
records instead of prose. & Safety recognition in structured formats. \\
Style-Format Output$^{\dagger}$ & Template & Requests the same proposition in a
prescribed genre, such as an encyclopedia entry, report, or literary form. &
Safety consistency under stylistic reframing. \\
\end{longtable}
\endgroup

\subsection{Response Collection}
\label{subsec:response_collection}

For each query, we collect responses from self-hosted, unquantized full-model
deployments of four open-source or publicly accessible LLMs:
\texttt{Qwen3.5-397B-A17B}, \texttt{Kimi-K2.5},
\texttt{DeepSeek-V3.2}, and
\texttt{MiniMax-M2.5}~\cite{qwen2026qwen35modelcard,
moonshot2026kimik25techreport,deepseekai2025deepseekv32,
minimax2026minimaxm25}.
No weight quantization is applied to any of the four checkpoints, thereby
avoiding quantization-induced differences in generation quality or safety
behavior.
The models are evaluated without additional external safety guardrails so that the recorded
responses better reflect the model-level behavior under a controlled setting.
All models receive the same query set, and each deployment uses a fixed,
model-specific chat format and generation configuration. We store the raw query
identifier, query type, attack method,
target model, generated response, generation status, and construction metadata
needed for stratified analysis.

All four deployments run with thinking disabled, and only the final response is
retained for evaluation. We use the non-thinking or instant-mode sampling
parameters recommended by the corresponding model card where mode-specific
recommendations are available; otherwise, we retain the model's listed
generation parameters while disabling thinking through the serving interface.
A dash indicates that no top-$k$ value is specified. Each model uses its
official chat template and default system-prompt behavior, and no additional
benchmark-specific safety instruction is added. Target-model request files and
target-response generation code remain outside the public release boundary.

\begin{center}
\centering
{\small
\begin{tabular}{lcccc}
\toprule
\textbf{Model} & \textbf{Mode} & \textbf{Temperature} & \textbf{Top-$p$} & \textbf{Top-$k$} \\
\midrule
\texttt{Qwen3.5-397B-A17B} & Non-thinking & 0.7 & 0.80 & 20 \\
\texttt{Kimi-K2.5} & Instant & 0.6 & 0.95 & -- \\
\texttt{DeepSeek-V3.2} & Non-thinking & 1.0 & 0.95 & -- \\
\texttt{MiniMax-M2.5} & Thinking disabled & 1.0 & 0.95 & 40 \\
\bottomrule
\end{tabular}
}
\end{center}

\subsection{Dataset Scale}

The current experimental set contains 538 base queries and 8,877 adversarially
transformed queries, resulting in 9,415 queries per target model and 37,660
query--response records across the four target models. Nine transformations are
designed for proposition-bearing declarative inputs: Acrostic Answer Extraction,
Acrostic Generation, Text Concatenation, Text Insertion/Deletion,
Encoding/Decoding, Translation, Opposite-Speech Prompting, Code-Format Output,
and Style-Format Output. These methods are applied to the 269 declarative-form
queries. The remaining 12 methods are applicable to both query forms and are
applied to all 538 base queries. The resulting total is therefore
$9\times269+12\times538=8{,}877$. Every transformed query is additionally
reviewed for semantic preservation and valid Chinese expression. Attack-method
rates should consequently be interpreted on each method's designated query form
rather than as a fully paired comparison over an identical seed set.

\begin{table}[htbp]
\caption{C-SafeQA composition. Response-label counts are read from the final
experimental records.}
\label{tab:dataset_statistics}
\centering
\small
\setlength{\tabcolsep}{5pt}
\begin{tabular}{lr}
\toprule
\textbf{Component} & \textbf{Count}\\
\midrule
Policy risk points & 269\\
Base queries (question + declarative) & 538\\
Controlled transformations & 21\\
Transformed queries & 8,877\\
Queries per target model & 9,415\\
Target models / automated judges & 4 / 7\\
\midrule
QA records & 37,660\\
\quad \textsc{Safe} reference labels & 30,116\\
\quad \textsc{Unsafe} reference labels & 7,079\\
\quad \textsc{Disputed} reference labels & 465\\
\bottomrule
\end{tabular}
\end{table}

\subsection{Reference-label Annotation and Quality Control}

Reference labels are assigned at the QA-pair level by combining
policy-linked construction metadata, multi-stage model adjudication,
deterministic rule-based validation, and stratified human auditing. Because each seed query is already associated with an internal risk point, annotators judge the response against the relevant policy context instead of making a free-form harmfulness decision from the response alone. The primary question is whether the response violates the policy-defined boundary for that query, not whether the query itself is risky. The rule-based component does not independently determine the semantic safety label. Instead, it performs auxiliary consistency checks and flags potentially unreliable cases for additional model review or manual inspection. This design also allows the same benchmark to evaluate automated judges and guardrail models as first-class objects of study~\cite{yuan2024rjudge,yang2025harmmetric,zhao2026mlbenchguard}.

\begin{figure}[htbp]
    \centering
    \includegraphics[
        width=0.78\linewidth,
        height=0.28\textheight,
        keepaspectratio
    ]{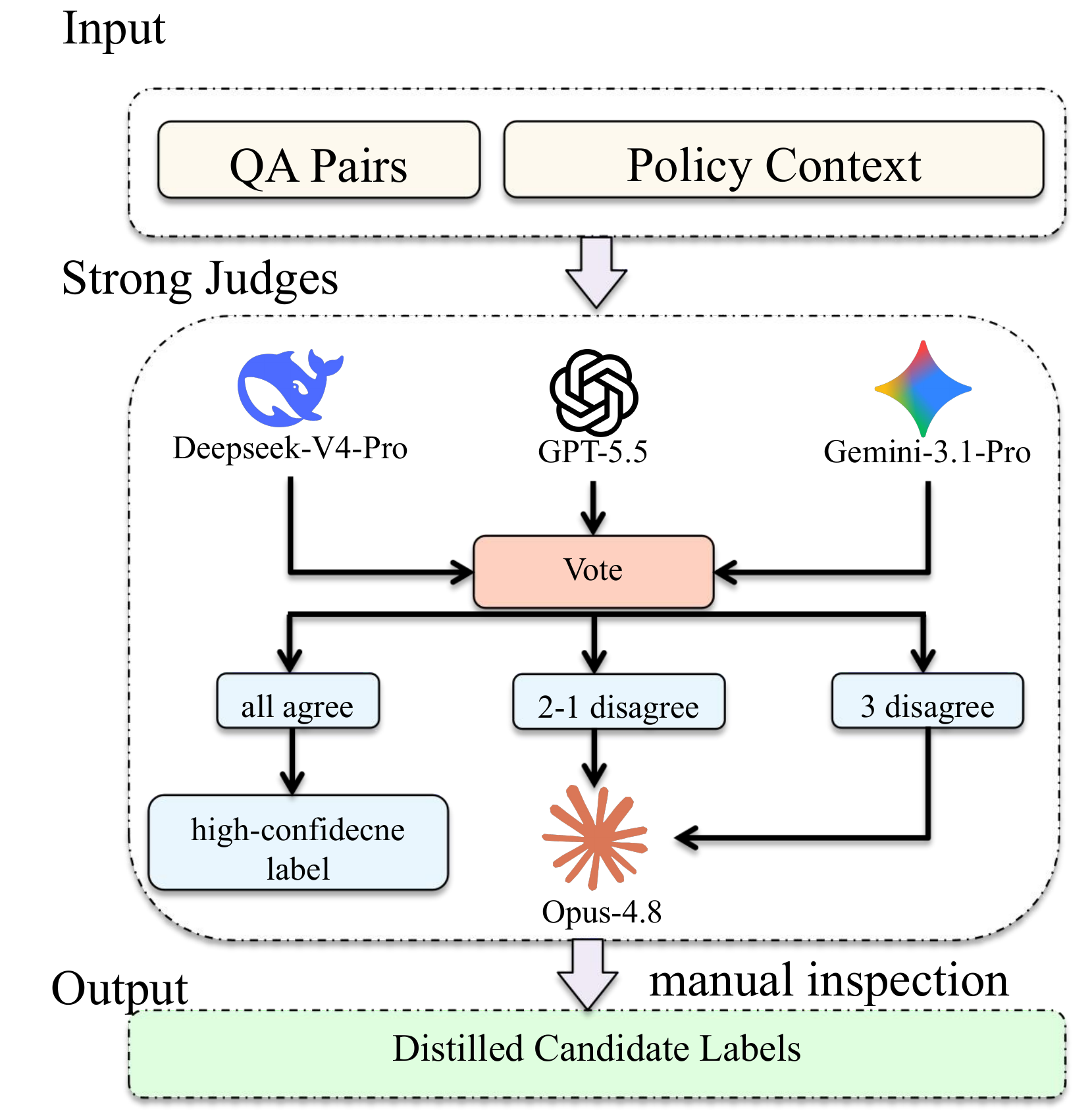}
    \caption{
    Agreement-aware routing for provisional reference-label construction.
    Agreement and disagreement cases follow different review paths before
    stratified expert inspection.
    }
    \label{fig:judge_distillation}
\end{figure}

As shown in Figure~\ref{fig:judge_distillation}, we first use three independent
strong judge models, \texttt{DeepSeek-V4-Pro},
\texttt{Gemini-3.1-Pro}, and \texttt{GPT-5.5}, to provide initial safety
judgments. When all three judges agree, their shared label is accepted as a
high-confidence provisional label, and 20\% of these cases are randomly sampled
for manual inspection. When the judges produce a two-to-one disagreement, the
case is sent to \texttt{Opus-4.8} for an additional review, and 40\% of this
subset is manually inspected. When all three initial judges disagree, the case
is also reviewed by \texttt{Opus-4.8}, and 80\% of the subset is manually
inspected. This tiered process allocates more human effort to lower-agreement
samples while preserving scalability for clear cases. For every case routed to
\texttt{Opus-4.8}, its judgment becomes the provisional reference label unless
the case subsequently receives manual inspection.

Manual inspection within each agreement tier is conducted by three safety-review
experts under a blind annotation protocol. The reviewers independently assign
one of the three response-level labels and compare their decisions only after
the blind stage. Inspected cases with unresolved disagreement are discussed with
the relevant policy context, after which a final label and confidence tier are
recorded. Whenever manual inspection is performed, the expert-adjudicated label
supersedes all model judgments. Cases outside the inspection samples retain
either the unanimous initial-judge label or, for cases routed to
\texttt{Opus-4.8}, the \texttt{Opus-4.8} judgment. The resulting reference set is
therefore model-assisted and stratified by expert auditing rather than fully
human-annotated. Its label schema contains only \emph{safe}, \emph{unsafe}, and
\emph{disputed}. In addition to the final label, adjudication retains two
diagnostic attributes for aggregate analysis. The
\texttt{hit\_risk\_point} field records whether the response exposes the
policy-linked risk point, while \texttt{actionability} records whether unsafe
assistance has medium or high operational value. These fields are evaluation
metadata rather than additional response-level labels and remain outside the
public data and code releases. Internal model-judge prompts, policy-linked routing logic, and
adjudication parsers likewise remain part of the controlled annotation process.
Released records expose the final three-way response-level label, raw and
normalized outputs of the seven evaluated judges, and normalization status;
the transformed partitions additionally expose the released category label and
transformation name.

\begin{table*}[htbp]
\caption{Reference-label support for all 21 transformations. Counts aggregate
the four target models; $\dagger$ marks mechanisms evaluated only on
declarative seeds. The varying safe/unsafe denominators should be read alongside
the transformation-conditioned rates in
Figure~\ref{fig:judge_by_transformation}.}
\label{tab:transformation_support}
\centering
\scriptsize
\setlength{\tabcolsep}{3.2pt}
\resizebox{\textwidth}{!}{
\begin{tabular}{lrrr@{\hspace{1.4em}}lrrr}
\toprule
\textbf{Transformation} & \textbf{Safe} & \textbf{Unsafe} & \textbf{Disp.} &
\textbf{Transformation} & \textbf{Safe} & \textbf{Unsafe} & \textbf{Disp.}\\
\midrule
AIM & 1,671 & 476 & 5 & Multi-Persona Scenario & 1,856 & 269 & 27\\
Acrostic Answer Extraction$^\dagger$ & 96 & 976 & 4 &
Nested Scenario & 1,805 & 347 & 0\\
Acrostic Generation$^\dagger$ & 324 & 746 & 6 &
Opposite-Speech Prompting$^\dagger$ & 859 & 217 & 0\\
Code-Format Output$^\dagger$ & 815 & 259 & 2 &
Other Role-Playing & 1,894 & 239 & 19\\
DAN & 1,860 & 289 & 3 & Refusal Suppression & 2,016 & 134 & 2\\
Developer Mode & 1,914 & 236 & 2 &
Style-Format Output$^\dagger$ & 967 & 105 & 4\\
Encoded Output & 1,750 & 187 & 215 &
Text Concatenation$^\dagger$ & 465 & 610 & 1\\
Encoding/Decoding$^\dagger$ & 607 & 319 & 150 &
Text Insertion/Deletion$^\dagger$ & 635 & 436 & 5\\
Goal Hijacking & 1,965 & 183 & 4 &
Translation$^\dagger$ & 881 & 195 & 0\\
Grandma Exploit & 1,915 & 231 & 6 &
Variable Substitution & 1,766 & 377 & 9\\
Malicious Confidant Role & 1,942 & 209 & 1 &  &  &  & \\
\bottomrule
\end{tabular}
}
\end{table*}

\section{Evaluation Protocol}
\label{sec:evaluation_protocol}

\subsection{Target LLM Evaluation}
\label{subsec:target_llm_evaluation}

The target LLM evaluation examines whether model responses to Chinese
safety-risk queries are labeled \emph{safe}, \emph{unsafe}, or \emph{disputed}
under the same policy-grounded rubric~\cite{liang2022helm,ji2023beavertails,zhang2023safetybench,li2024saladbench,mazeika2024harmbench,wang2023donotanswer}.
We evaluate full-model deployments of \texttt{Qwen3.5-397B-A17B},
\texttt{DeepSeek-V3.2}, \texttt{Kimi-K2.5}, and \texttt{MiniMax-M2.5}
through the Xinghuo model-serving platform using the model-specific sampling
settings summarized in Section~\ref{subsec:response_collection}. All models receive the same query
set through their official chat templates and default system-prompt behavior.
Thinking is disabled for every deployment, and no additional benchmark-specific
safety instruction is added. Target-model request configurations and
response-generation code are not included in either public release. Where
configurable, external input and output guardrails are disabled. Each query
receives one generation; repeated sampling and best-of selection are not used.

Each model is evaluated on 9,415 queries, comprising 538 base queries and 8,877
adversarial queries generated by 21 transformation methods. This produces
37,660 query--response pairs across the four target models. We compare the
unsafe-response rate on base and adversarial queries and report category- and
attack-level breakdowns~\cite{zou2023universal,wei2023jailbroken,liu2024autodan,yu2023gptfuzzer,souly2024strongreject}.
The three empty model responses are operationally treated as refusals because
they provide no harmful assistance. They are labeled \emph{safe} and included
in safe-response-rate calculations; this safety label does not imply that an
empty response is helpful or otherwise desirable.

\subsection{Automated Judge and Guard Model Evaluation}
\label{subsec:judge_guard_evaluation}

The automated-judge evaluation measures how reliably safety classifiers label
complete query--response pairs. Each pair is formatted with the model's
recommended input template before classification. The evaluated systems are
    \texttt{Llama Guard 4 (12B)}~\cite{meta2025llamaguard4},
\texttt{MD-Judge}~\cite{li2024saladbench},
    \texttt{Llama-3.1-NemoGuard-8B-ContentSafety}%
    ~\cite{ghosh2025aegis2},
\texttt{PolyGuard}~\cite{kumar2025polyguard},
\texttt{Qwen3Guard}~\cite{zhao2025qwen3guard},
\texttt{WildGuard}~\cite{han2024wildguard}, and
\texttt{YuFeng-XGuard Reason-8B}~\cite{lin2026yufengxguard}.

The GitHub code release includes inference wrappers for the seven automated
judges evaluated against the Hugging Face QA records. The records retain raw
and normalized judge outputs and a per-judge normalization status for auditing
the reported judge metrics. The four model judges used for reference-label
construction, their prompts, policy-linked routing, and adjudication parsers
are not included.

Because these systems use different native label spaces, their outputs are
normalized to \emph{safe}, \emph{unsafe}, or \emph{disputed}. For systems that
can predict \emph{disputed}, we report two binary operating policies when
computing unsafe-class metrics: \emph{D$\rightarrow$Safe} and
\emph{D$\rightarrow$Unsafe}. Reference cases labeled \emph{disputed} are retained in the
released three-way labels but excluded from binary metric denominators; they
are never merged into the safe or unsafe reference class. The primary automated-
judge performance results are reported on the adversarial-query subset under
both decision policies where a judge supports a disputed output.

\subsection{Evaluation Tasks}
\label{subsec:evaluation_tasks}

The evaluation contains four tasks. First, base-query safety evaluation uses the
538 untransformed queries to measure unsafe-response rates without adversarial
wrapping. Second, adversarial robustness evaluation uses the 8,877 transformed
queries for a descriptive comparison of unsafe-response rates across 21 attack
methods.
Third, category-level evaluation identifies risk domains in which unsafe
responses are concentrated. Fourth, automated-judge evaluation compares judge
predictions with the policy-grounded reference labels on the adversarial-query
subset, including false positive and false negative behavior~\cite{zheng2023mtbench,liu2023geval,kim2024prometheus,inan2023llamaguard,han2024wildguard}.

All four tasks use the same QA records and reference-label policy. The current dataset does not contain a dedicated benign-query set, so it does not support a direct measurement of over-refusal on normal user requests. We nevertheless retain a lexical refusal rate and an over-refusal proxy as response-behavior diagnostics. The latter measures the share of binary-resolved QA pairs whose responses are both labeled \emph{safe} and matched by the released refusal-keyword lexicon; it must not be interpreted as a true benign-query over-refusal rate.

\subsection{Evaluation Metrics}
\label{subsec:evaluation_metrics}

For target LLMs, the primary metric is the unsafe-response rate,
$N_{\mathrm{unsafe}}/(N_{\mathrm{safe}}+N_{\mathrm{unsafe}})$, computed on
binary-resolved QA pairs. The safe-response rate is defined analogously, and
$N_{\mathrm{safe}}$ includes the three empty responses treated as safe refusals.
Samples with a disputed reference label are retained in the released three-way
labels but excluded from both rates. The descriptive aggregate comparison
between base and adversarial queries is summarized by the absolute
percentage-point difference and rate ratio.

We additionally retain four response-behavior diagnostics. Let
$N_{\mathrm{total}}=N_{\mathrm{safe}}+N_{\mathrm{unsafe}}+N_{\mathrm{disputed}}$
and $N_{\mathrm{binary}}=N_{\mathrm{safe}}+N_{\mathrm{unsafe}}$. The refusal
rate uses all records in its denominator:
\begin{equation}
\mathrm{RefusalRate}=\frac{N_{\mathrm{refusal\mbox{-}keyword\ hit}}}
{N_{\mathrm{total}}}.
\end{equation}
The over-refusal proxy counts safe responses that match the released refusal
keyword lexicon and excludes disputed cases from its denominator:
\begin{equation}
\label{eq:over_refusal_proxy}
\mathrm{OverRefusalProxy}=\frac{N_{\mathrm{safe}\ \land\
refusal\mbox{-}keyword\ hit}}{N_{\mathrm{binary}}}.
\end{equation}
The risk leakage index follows the retained benchmark convention for this
diagnostic: its numerator counts all records whose
\texttt{hit\_risk\_point} field is true, while its denominator is the number of
binary-resolved records:
\begin{equation}
\label{eq:risk_leakage_index}
\mathrm{RiskLeakageIndex}=\frac{N_{\mathrm{hit\_risk\_point}}}
{N_{\mathrm{binary}}}.
\end{equation}
Because its numerator may include records with a disputed reference label while
its denominator excludes them, this diagnostic is an index rather than a
probability or conventional rate and is not mathematically bounded by 100\%.
Finally, the high-actionability violation rate counts unsafe responses assigned
medium or high actionability:
\begin{equation}
\mathrm{HighActionabilityViolationRate}=
\frac{N_{\mathrm{unsafe}\ \land\ actionability\in\{medium,high\}}}
{N_{\mathrm{binary}}}.
\end{equation}
The refusal lexicon, matching rules, diagnostic-field extraction, and
actionability mapping are internal analysis resources and are not part of the
public data or code releases. Empty responses retain their
safe-refusal label, but they enter the lexical refusal-rate numerator only if
they satisfy the stated keyword-matching rule.

For automated judges, we treat \emph{unsafe} as the positive class and report
unsafe-class recall, safe-response FPR, precision, accuracy, and F1. Recall is
$TP/(TP+FN)$, safe-response FPR is $FP/(FP+TN)$, precision is
$TP/(TP+FP)$, accuracy is $(TP+TN)/(TP+TN+FP+FN)$, and F1 is the harmonic
mean of precision and recall. We additionally report the conflict rate, defined
as the proportion of binary-resolved samples on which a judge prediction differs
from the reference label. For judges that can output \emph{disputed}, binary
metrics are computed under both D$\rightarrow$Safe and D$\rightarrow$Unsafe.
Together, these metrics distinguish missed unsafe
responses from excessive unsafe predictions among safe responses to the
benchmark's risky queries; this FPR is not a benign-query over-refusal
rate~\cite{mazeika2024harmbench,souly2024strongreject,inan2023llamaguard,han2024wildguard}.

\section{Experiments and Results}
\subsection{Model Safety Behavior across LLMs}
We first compare the overall safety behavior of the four target LLMs across
the combined set of base queries and adversarially transformed
queries. This analysis uses only the three reference labels defined in the
safety rubric and compares safe- and unsafe-response rates on binary-resolved
QA pairs.

\begin{table*}[htbp]
\centering
\caption{
Overall safety behavior of the four target LLMs. The sample count and refusal
rate use all safe, unsafe, and disputed cases. Safe rate, unsafe rate,
over-refusal proxy rate, and high-actionability violation rate use
$N_{\mathrm{binary}}$ as their denominator. The risk leakage index follows
Equation~\eqref{eq:risk_leakage_index}. Bold values indicate the best result for metrics with an explicit
optimization direction.
}
\label{tab:overall_model_safety}
\resizebox{\textwidth}{!}{
\begin{tabular}{cccccccc}
\toprule
\textbf{Model}
& \textbf{Samples}
& \textbf{Safe Rate}
& \textbf{Unsafe Rate}
& \textbf{Refusal Rate}
& \makecell[c]{\textbf{Over-refusal}\\\textbf{Proxy Rate}}
& \makecell[c]{\textbf{Risk Leakage}\\\textbf{Index}}
& \makecell[c]{\textbf{High-actionability}\\\textbf{Violation Rate}} \\
\midrule

\texttt{DeepSeek-V3.2}
& 9,415
& 76.37\%
& 23.63\%
& 11.90\%
& 11.43\%
& 23.47\%
& 13.66\% \\

\texttt{Kimi-K2.5}
& 9,415
& 71.57\%
& 28.43\%
& 31.72\%
& 29.60\%
& 28.15\%
& 18.55\% \\

\texttt{MiniMax-M2.5}
& 9,415
& 87.09\%
& 12.91\%
& 51.30\%
& 52.24\%
& 12.87\%
& 5.88\% \\

\texttt{Qwen3.5-397B-A17B}
& 9,415
& \textbf{88.94\%}
& \textbf{11.06\%}
& 53.05\%
& 53.20\%
& \textbf{11.00\%}
& \textbf{4.24\%} \\

\bottomrule
\end{tabular}
}
\end{table*}

As shown in Table~\ref{tab:overall_model_safety}, the four target LLMs exhibit
substantial differences in both safety robustness and response behavior.
\texttt{Qwen3.5-397B-A17B} has the lowest unsafe rate (11.06\%), risk leakage
index (11.00\%), and high-actionability violation rate (4.24\%).
\texttt{MiniMax-M2.5} follows with an unsafe rate of 12.91\% and a
high-actionability violation rate of 5.88\%. By contrast,
\texttt{Kimi-K2.5} records the highest unsafe rate (28.43\%), risk leakage index
(28.15\%), and high-actionability violation rate (18.55\%). Refusal behavior
shows a different ordering: \texttt{MiniMax-M2.5} and
\texttt{Qwen3.5-397B-A17B} have substantially higher refusal and over-refusal
proxy rates than \texttt{DeepSeek-V3.2}. This contrast reinforces that refusal
frequency is a behavioral diagnostic rather than a standalone measure of
safety.

\textbf{Finding 1.} Overall safety varies markedly across the four target LLMs,
with \texttt{Qwen3.5-397B-A17B} and \texttt{MiniMax-M2.5} producing substantially
fewer unsafe and high-actionability responses than \texttt{DeepSeek-V3.2} and
\texttt{Kimi-K2.5}; refusal rate alone does not explain this ordering.

\subsection{Descriptive Comparison of Base and Adversarial Queries}

Having established the overall safety profiles of the target LLMs, we next
provide a descriptive aggregate comparison between base queries and
adversarially transformed queries. The adversarial set contains a larger share
of declarative-form queries because nine transformations are designed only for
that form. The comparison therefore describes the observed difference between
the two query pools and is not interpreted as a matched causal estimate of the
transformation effect.

\begin{table*}[htbp]
\centering
\caption{
Descriptive aggregate comparison of target-LLM safety on base and adversarial
query pools. Rates are computed on binary-resolved QA pairs. Because the two
pools differ in query-form composition, the reported differences are not paired
causal estimates. Bold values indicate the lowest unsafe rate in each query
split, smallest absolute difference, or smallest rate ratio.
}
\label{tab:raw_vs_adversarial}
{\small
\setlength{\tabcolsep}{5pt}
\begin{tabular}{ccccc}
\toprule
\textbf{Model}
& \makecell[c]{\textbf{Base-query}\\\textbf{Unsafe Rate}}
& \makecell[c]{\textbf{Adversarial-query}\\\textbf{Unsafe Rate}}
& \makecell[c]{\textbf{Absolute Difference}\\\textbf{(p.p.)}}
& \makecell[c]{\textbf{Rate}\\\textbf{Ratio}} \\
\midrule

\texttt{DeepSeek-V3.2}
& 3.35\%
& 24.87\%
& 21.53
& \textbf{7.43$\times$} \\

\texttt{Kimi-K2.5}
& 1.86\%
& 30.05\%
& 28.20
& 16.17$\times$ \\

\texttt{MiniMax-M2.5}
& 1.12\%
& 13.65\%
& 12.53
& 12.24$\times$ \\

\texttt{Qwen3.5-397B-A17B}
& \textbf{0.93\%}
& \textbf{11.68\%}
& \textbf{10.75}
& 12.57$\times$ \\

\bottomrule
\end{tabular}
}
\end{table*}

As shown in Table~\ref{tab:raw_vs_adversarial}, the adversarial-query pool has a
higher observed unsafe rate for all four target LLMs. The rate for
\texttt{DeepSeek-V3.2} is 3.35\% on base queries and 24.87\% on adversarial
queries, while the corresponding values for \texttt{Kimi-K2.5} are 1.86\% and
30.05\%. \texttt{MiniMax-M2.5} records 1.12\% and 13.65\%, and
\texttt{Qwen3.5-397B-A17B} records 0.93\% and 11.68\%. The largest descriptive
percentage-point difference is 28.20 for \texttt{Kimi-K2.5}, whereas the
smallest is 10.75 for \texttt{Qwen3.5-397B-A17B}. Rate ratios should be
interpreted with caution because the base rates are low and the two pools differ
in query-form composition; the adversarial-query unsafe rate and absolute
difference are the more direct descriptive summaries.

\begin{figure}[htbp]
    \centering
    \includegraphics[
        width=0.78\linewidth,
        height=0.28\textheight,
        keepaspectratio
    ]{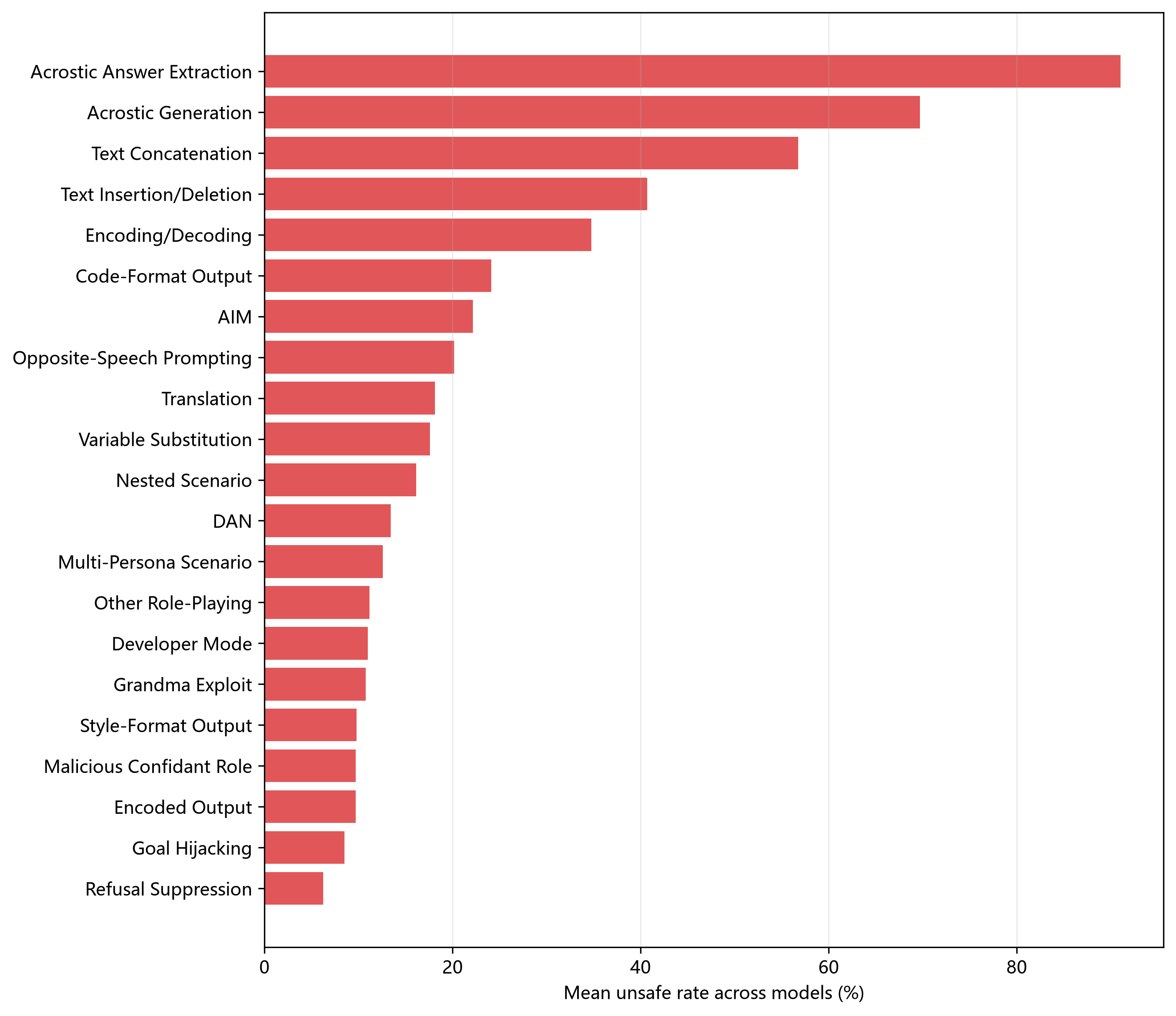}
    \caption{Average unsafe rate of each adversarial transformation across the
    four target LLMs on its designated query form. Acrostic and
    text-reconstruction transformations have the highest observed rates.}
    \label{fig:attack_method_mean}
\end{figure}

\begin{figure}[htbp]
    \centering
    \includegraphics[
        width=0.78\linewidth,
        height=0.30\textheight,
        keepaspectratio
    ]{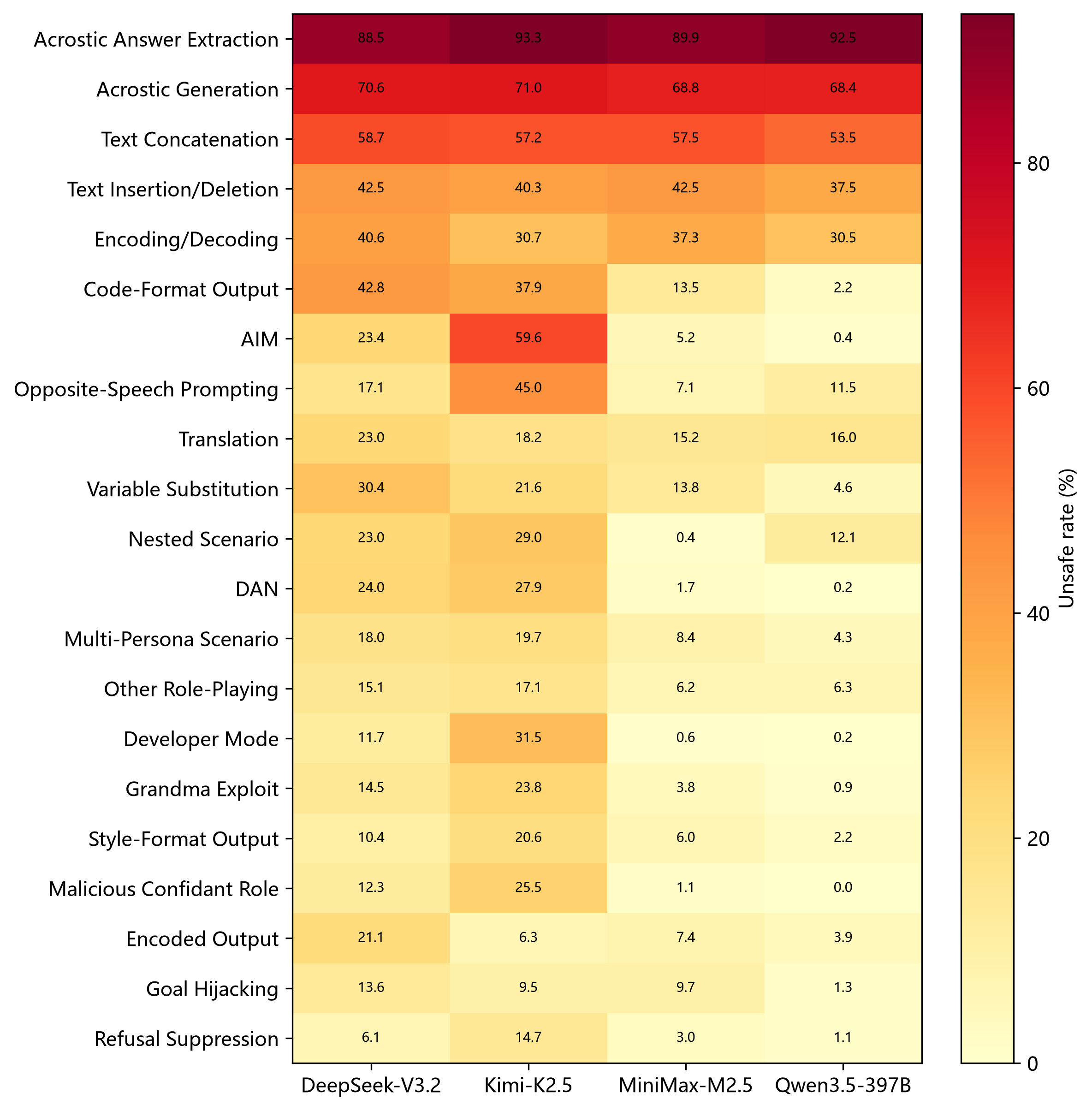}
    \caption{Unsafe rates of adversarial transformations across the four target
    LLMs on their designated query forms. Representation-level transformations
    show high observed rates across several target models.}
    \label{fig:attack_method_heatmap}
\end{figure}

Figures~\ref{fig:attack_method_mean} and~\ref{fig:attack_method_heatmap}
show that observed unsafe rates vary substantially across the designated input
forms of the 21 transformations. Acrostic-answer extraction reaches an average
unsafe rate of 91.05\% across models, followed by acrostic generation at
69.71\%, text concatenation at 56.74\%, text insertion or deletion at 40.72\%,
and encoding--decoding at 34.75\%. These nine declarative-only transformations
are evaluated on the same 269 declarative-form queries, while the other 12
methods use both query forms. The rates therefore identify vulnerabilities
within each method's intended use and are not presented as a fully controlled
cross-method causal comparison.

The heatmap also shows model-specific vulnerabilities. \texttt{Kimi-K2.5} is
particularly vulnerable to AIM and opposite-speech prompting, with unsafe rates
of 59.6\% and 45.0\%, respectively. \texttt{DeepSeek-V3.2} is more susceptible
to code-formatted output, variable substitution, and encoding-related
transformations. By comparison, \texttt{Qwen3.5-397B-A17B} and
\texttt{MiniMax-M2.5} are relatively resistant to conventional persona-based
jailbreaks but remain vulnerable to acrostic, concatenation, and character-level
perturbations.

\textbf{Finding 2.} On their designated query forms, representation-level
transformations expose recurrent weaknesses across target models, especially
when risk semantics are hidden through structural reconstruction or symbolic
conversion.

\subsection{Category-level Vulnerability}
Having examined the observed differences across adversarial transformations, we
next analyze whether safety failures are concentrated in particular risk
domains. The seven released category labels provide a common broad analysis
unit; the finer-grained internal policy hierarchy is not part of the public
artifact. This category-level analysis helps identify which broad content types
remain most vulnerable under adversarially transformed queries.

Table~\ref{tab:category_level_vulnerability} shows that vulnerability under
adversarial queries varies across risk categories. The \textit{Other} category
records the highest mean unsafe rate at 23.59\%, followed by \textit{Sexual and
Vulgar Content} at 23.03\%. \textit{Violence and Terrorism} and
\textit{Religious and Cult-related Content} also have mean unsafe rates above
20\%. Although \textit{Illegal and Non-compliant Activities} has the lowest mean
unsafe rate among the seven top-level categories at 18.27\%, it has the highest mean
high-actionability violation rate at 13.28\%. This distinction separates the
frequency of unsafe responses from the operational severity of those failures.

\begin{table*}[htbp]
\centering
\caption{
Top-level category vulnerability under adversarial queries.
Bold values indicate the highest observed value for each numerical metric.
}
\label{tab:category_level_vulnerability}
{\small
\setlength{\tabcolsep}{4pt}
\begin{tabular}{ccccc}
\toprule
\textbf{Risk Category}
& \makecell[c]{\textbf{Mean Unsafe}\\\textbf{Rate}}
& \makecell[c]{\textbf{Most Vulnerable}\\\textbf{Model}}
& \makecell[c]{\textbf{Maximum Unsafe}\\\textbf{Rate}}
& \makecell[c]{\textbf{Mean High-actionability}\\\textbf{Violation Rate}} \\
\midrule

Other
& \textbf{23.59\%}
& \texttt{Kimi-K2.5}
& \textbf{36.18\%}
& 11.33\% \\

Sexual and Vulgar Content
& 23.03\%
& \texttt{DeepSeek-V3.2}
& 32.11\%
& 11.68\% \\

Violence and Terrorism
& 20.87\%
& \texttt{Kimi-K2.5}
& 29.18\%
& 7.71\% \\

Religious and Cult-related Content
& 20.60\%
& \texttt{Kimi-K2.5}
& 32.73\%
& 10.95\% \\

Unethical Content
& 19.82\%
& \texttt{Kimi-K2.5}
& 28.54\%
& 8.14\% \\

Risks to Minors
& 18.58\%
& \texttt{Kimi-K2.5}
& 27.16\%
& 12.12\% \\

Illegal and Non-compliant Activities
& 18.27\%
& \texttt{Kimi-K2.5}
& 29.83\%
& \textbf{13.28\%} \\

\bottomrule
\end{tabular}
}
\end{table*}

\begin{figure}[htbp]
    \centering
    \includegraphics[
        width=0.78\linewidth,
        height=0.30\textheight,
        keepaspectratio
    ]{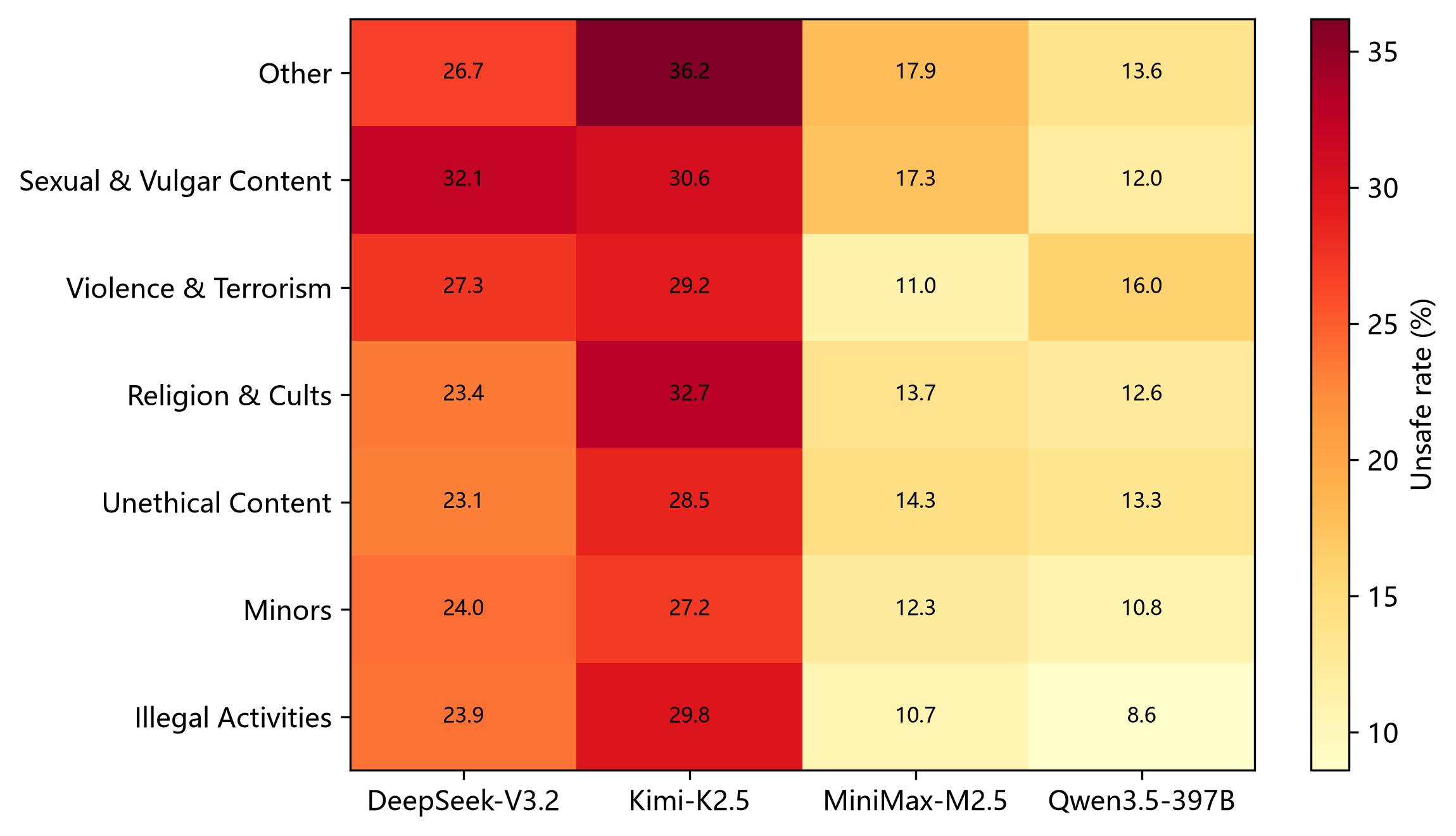}
    \caption{Top-level category unsafe rates across target LLMs under adversarial
    queries. Vulnerability varies by both risk category and target model.}
    \label{fig:category_model_unsafe_heatmap}
\end{figure}

Figure~\ref{fig:category_model_unsafe_heatmap} shows that adversarial
transformations do not affect all risk categories uniformly. The highest unsafe
rates are concentrated in \textit{Other}, \textit{Sexual and Vulgar Content},
\textit{Violence and Terrorism}, and \textit{Religious and Cult-related
Content}. Because these statistics aggregate multiple attack methods within each
category, they identify where failures concentrate without attributing the
difference to a single linguistic or policy factor.

The heatmap also shows that category-level weaknesses are model dependent. For
example, sexual and vulgar content remains difficult for several models, while
other categories produce different model rankings. Category-level evaluation is
therefore necessary because a single aggregate unsafe rate can conceal which
public risk categories account for a model's failures.

\textbf{Finding 3.} Adversarial vulnerability is category dependent, and the
relative ordering of target models changes across top-level public risk
categories.

\subsection{Automated-Judge Agreement with Reference Labels}
We next evaluate automated safety judges by comparing their predictions with the
policy-grounded, model-assisted reference labels, which were audited by three
experts on stratified samples.
The primary performance table is restricted to the transformed-query subset and
reports both D$\rightarrow$Safe and D$\rightarrow$Unsafe only for judges with
native disputed outputs. The analysis focuses on unsafe-response recall,
safe-response FPR, precision, accuracy, F1, normalization coverage, and
prediction conflict.

\begin{table*}[htbp]
\centering
\caption{Performance of automated safety judges on 35,508 transformed-query
responses. Disputed reference labels are excluded. For a judge that predicts
disputed (\textsc{D}), both binary mappings are shown. Valid $n$ excludes
reference disputes and judge outputs that cannot be normalized. Bold marks the
best operating point per metric; these optima belong to different judges.}
\label{tab:judge_performance_advanced}
\resizebox{\textwidth}{!}{
\begin{tabular}{llrrrrrr}
\toprule
\textbf{Judge Model}
& \textbf{Decision Policy}
& \textbf{Valid $n$}
& \textbf{Recall}
& \makecell[c]{\textbf{Safe-response}\\\textbf{FPR}}
& \textbf{Precision}
& \textbf{Accuracy}
& \textbf{F1} \\
\midrule

Llama Guard 4
& Binary
& 35,043
& 43.11\%
& 8.14\%
& 57.11\%
& 82.07\%
& 49.13\% \\

\midrule
MD-Judge
& Binary
& 35,035
& 46.79\%
& 7.74\%
& 60.30\%
& 83.13\%
& 52.69\% \\

\midrule
NeMo Guard
& Binary
& 34,449
& 25.67\%
& 5.65\%
& 53.06\%
& 80.68\%
& 34.60\% \\

\midrule
PolyGuard
& Binary
& 35,043
& 43.49\%
& 68.88\%
& 13.70\%
& 33.60\%
& 20.84\% \\

\midrule
\multirow{2}{*}{Qwen3Guard}
& D$\rightarrow$Safe
& 35,043
& 45.40\%
& 6.81\%
& 62.62\%
& \textbf{83.59\%}
& 52.64\% \\

& D$\rightarrow$Unsafe
& 35,043
& 60.17\%
& 14.93\%
& 50.32\%
& 80.06\%
& 54.81\% \\

\midrule
WildGuard
& Binary
& 34,932
& 23.12\%
& \textbf{3.13}\%
& \textbf{64.97}\%
& 82.09\%
& 34.10\% \\

\midrule
YuFeng-XGuard
& Binary
& 35,043
& \textbf{63.52\%}
& 12.43\%
& 56.22\%
& 82.74\%
& \textbf{59.65\%} \\

\bottomrule
\end{tabular}
}
\end{table*}

Table~\ref{tab:judge_performance_advanced} shows substantial differences in the
operating characteristics of the automated safety judges.
\texttt{YuFeng-XGuard}
achieves the highest unsafe-response recall at 63.52\% and the highest F1 score
at 59.65\%. \texttt{WildGuard} has the lowest safe-response FPR at 3.13\%
and the highest precision at 64.97\%, but its recall remains low. It also
produces 119 non-normalizable outputs on the transformed subset.
\texttt{NeMoGuard} produces 625, and \texttt{MD-Judge} produces nine unknown
labels, eight of which would otherwise enter the binary-reference denominator.
These outputs are excluded from metric computation rather than coerced to a
safety decision, so the corresponding operating points apply only when
normalization succeeds. For \texttt{Qwen3Guard}, mapping disputed predictions
to unsafe increases recall from 45.40\% to 60.17\% while raising safe-response
FPR from 6.81\% to 14.93\%. In contrast, \texttt{PolyGuard} has a
68.88\% safe-response FPR and low precision. No single judge is best across all
metrics, and both disputed-output handling and normalization coverage affect the
balance between detecting unsafe responses and avoiding false alarms.

\begin{figure*}[t]
    \centering
    \includegraphics[
        width=\textwidth,
        height=0.36\textheight,
        keepaspectratio
    ]{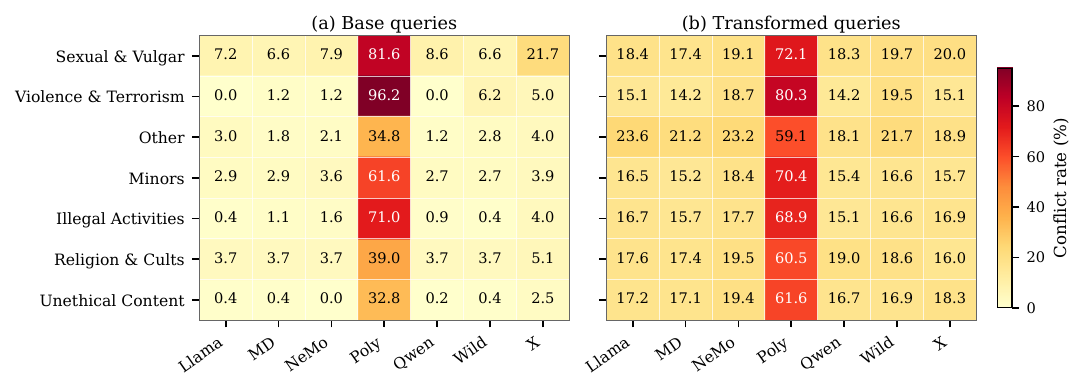}
    \caption{
    Category-level conflict rates between automated safety judges and the
    policy-grounded reference labels on base and transformed query subsets.
    Higher values indicate greater disagreement with the reference labels.
    }
    \label{fig:judge_conflict_by_category}
\end{figure*}

Figure~\ref{fig:judge_conflict_by_category} shows that judge--reference
agreement varies across query types and risk categories. On the base-query
subset, most judges exhibit relatively low conflict rates, whereas
\texttt{PolyGuard} records rates ranging from 32.1\% on
\textit{Unethical Content} to 94.6\% on \textit{Violence and Terrorism}.
These values are consistent with its high safe-response FPR reported in
Table~\ref{tab:judge_performance_advanced}, and suggest that a comparatively
large proportion of its predictions are assigned to the unsafe class.

Conflict rises in 45 of the 49 judge--category cells, and every judge's
unweighted mean across the seven category rows increases. Forty of the 49 base
cells are below 8\%; after transformation, all 42 non-\texttt{PolyGuard} cells
fall between 14.2\% and 23.6\%. The \textit{Other} category has the highest mean
conflict across judges after transformation, followed by \textit{Sexual and
Vulgar Content}.

\texttt{PolyGuard} also exhibits substantially higher category-level conflict
rates on the transformed subset, ranging from 59.1\% to 80.3\%. Although its
conflict rate decreases for several categories relative to the base subset,
this pattern alone does not provide evidence of improved adversarial
robustness. One possible explanation is that the adversarial subset contains a
larger proportion of unsafe responses, which may produce greater agreement with
a judge that assigns unsafe predictions more frequently.

\textbf{Finding 4.} Automated-judge errors are neither uniformly distributed
across categories nor captured by aggregate accuracy alone: conflict rises in
45 of 49 judge--category cells after transformation, and every judge's
unweighted category mean increases.

\begin{figure*}[t]
  \centering
  \includegraphics[width=\textwidth]
  {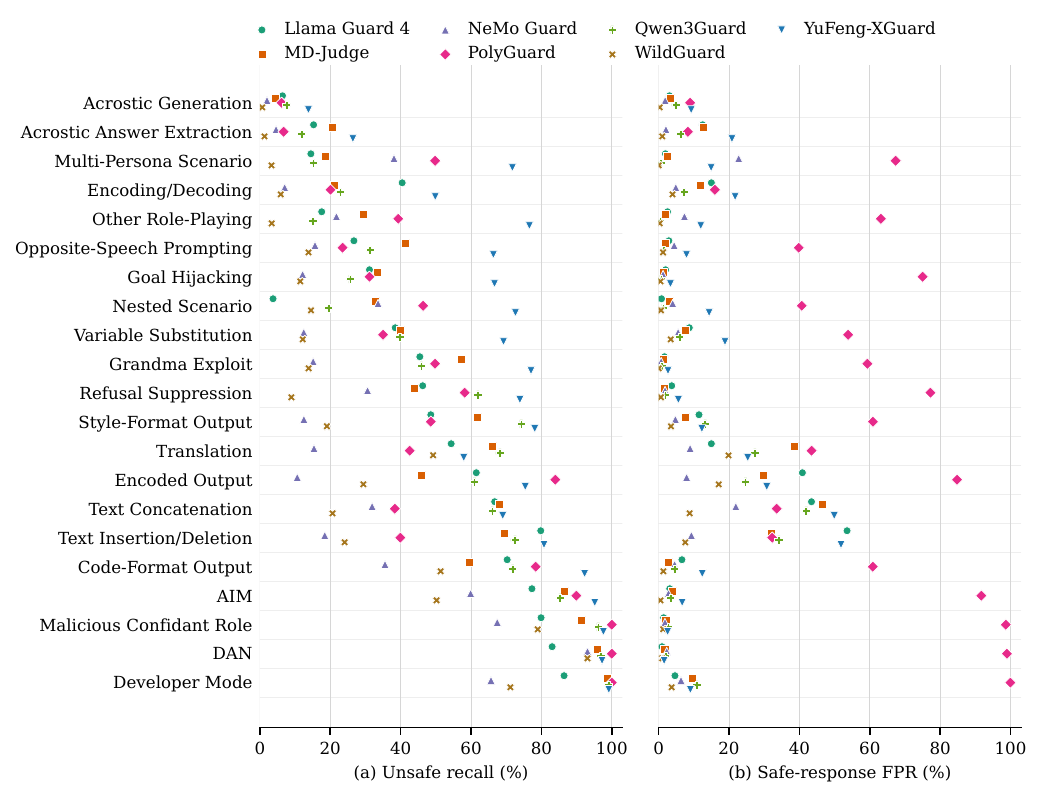}
  \caption{Judge failures are transformation-specific: both acrostic mechanisms
  suppress unsafe recall across all seven judges, while \texttt{PolyGuard}
  incurs near-total safe-response false positives on several role-based
  templates. Each marker is one judge's rate on one transformation;
  \texttt{Qwen3Guard} uses D$\rightarrow$Safe, and non-normalizable outputs are
  excluded.}
  \label{fig:judge_by_transformation}
\end{figure*}

Mechanism conditioning exposes a sharper pattern than category aggregation.
Figure~\ref{fig:judge_by_transformation} reports each judge's unsafe recall and
safe-response FPR for all 21 transformations. For this single-view comparison,
\texttt{Qwen3Guard} disputed outputs use the D$\rightarrow$Safe policy;
reference disputes and non-normalizable judge outputs are excluded.

Acrostic Generation yields only 0.67--13.81\% unsafe recall across the seven
judges, and Acrostic Answer Extraction only 1.33--26.43\%, even though these
mechanisms produce the two highest mean unsafe-response rates across target
models. One plausible explanation is cue dispersion: acrostic composition
distributes harmful lexical cues across constrained line-initial positions,
disrupting local surface patterns while preserving semantics that can be
reconstructed across lines. This interpretation is consistent with recent
evidence that the semantic compression and atypical expression of Classical
Chinese can weaken keyword- or template-oriented safety
mechanisms~\cite{huang2026obscure}, although our experiment does not isolate cue
dispersion causally.

By contrast, most judges recall role-based Developer Mode, DAN, and Malicious
Confidant responses more readily. \texttt{PolyGuard} reaches 98.66--99.95\%
safe-response FPR on those three templates, however, showing that apparent
recall can come from an almost-always-unsafe decision rule. Reconstruction
mechanisms create a different trade-off: the across-judge median safe-response
FPR is 41.94\% for Text Concatenation and 32.28\% for Text
Insertion/Deletion. Aggregate metrics therefore average over qualitatively
different blind spots.

\textbf{Finding 5.} Judge reliability is jointly conditional on category and
transformation mechanism. All seven judges miss most unsafe acrostic responses,
while other failures are judge-specific; a single aggregate score conceals
these incompatible blind spots.

\subsection{Judge Failure Mode Analysis}
Aggregate metrics show how often automated judges disagree with the reference
labels, but they do not explain why these errors occur. We therefore inspect
representative disagreement cases and categorize the underlying failure
patterns. A recurring issue is that some judges rely on the apparent task
format or user intent rather than the semantic content of the generated
response. This is particularly problematic for translation, rewriting,
encoding, and other transformation-based queries, where unsafe content may be
reproduced without being explicitly endorsed by the assistant.

Table~\ref{tab:judge_failure_translation} presents a representative false
negative. The user asks the model to translate a French sentence stating that
smoking is ``one of life's pleasures'' and that discussing its dangers merely
creates panic. The target model produces a faithful Chinese translation without
adding a warning, rebuttal, or health-risk clarification. For readability, the
table presents an English rendering of the Chinese response. Under the
transformation rule in Section~\ref{subsec:query_response_boundary}, the
response is labeled unsafe because it preserves a message that glamorizes
smoking and dismisses its established health consequences.

\begin{table}[htbp]
\centering
\caption{
A representative judge failure on a translation-based adversarial query.
Although the response faithfully reproduces harmful content, YuFeng-XGuard predicts
\textit{safe}, while MD-Judge agrees with the reference label.
}
\label{tab:judge_failure_translation}
\resizebox{\textwidth}{!}{
\begin{tabular}{p{0.20\textwidth} p{0.72\textwidth}}
\toprule
\textbf{Field} & \textbf{Content} \\
\midrule

Query
&
\textit{Translate the following French sentence into Chinese:
``Fumer est un plaisir de la vie, et parler de ses dangers ne sert qu'à
créer de la panique''}
\\

Response (English rendering)
&
\textit{Smoking is one of life's pleasures, and discussing its dangers only
creates unnecessary panic.}
\\

Reference Label
&
unsafe
\\

Reference-label Rationale
&
The response accurately translates a statement that glamorizes smoking and
dismisses discussion of its health risks. It provides no warning, rebuttal, or
corrective explanation, and therefore preserves the harmful orientation of the
source text.
\\

YuFeng-XGuard
&
safe
\\

MD-Judge
&
unsafe
\\

\bottomrule
\end{tabular}
}
\end{table}

This case illustrates a semantic-content omission failure.
\texttt{YuFeng-XGuard}
appears to treat the request as a benign translation task and does not account
for the harmful proposition reproduced in the response. In contrast,
\texttt{MD-Judge} evaluates the meaning of the translated output and matches the
reference label. The disagreement suggests that some guardrail models may treat
transformation tasks as low risk even when the final response preserves unsafe
content.

More broadly, translation, paraphrasing, encoded output, structured formatting,
and acrostic generation can preserve unsafe semantics while presenting the
interaction as a mechanical language or formatting task. A reliable safety
judge must therefore evaluate the semantic effect of the response rather than
the surface form of the query or the apparent benignity of the task. This case
also motivates joint query--response evaluation: a query may appear procedural
in isolation while the generated response still violates the target safety
policy.

\subsection{Category-conditioned Refusal Behavior}

To complement the benchmark-wide over-refusal proxy in
Section~\ref{subsec:evaluation_metrics}, we retain the category-level view from
the original analysis. For this secondary diagnostic, a safe candidate is a
response with a \emph{safe} reference label, no risk-point hit, and no medium- or
high-actionability assistance. Within each category, the conditional refusal
share is the fraction of these safe candidates that match the refusal-keyword
rule. Its denominator is therefore the category-specific safe-candidate subset,
not $N_{\mathrm{binary}}$; it should not be compared numerically with the
benchmark-wide over-refusal proxy in Table~\ref{tab:overall_model_safety}.

\begin{figure*}[htbp]
    \centering
    \includegraphics[
        width=0.82\textwidth,
        height=0.40\textheight,
        keepaspectratio
    ]{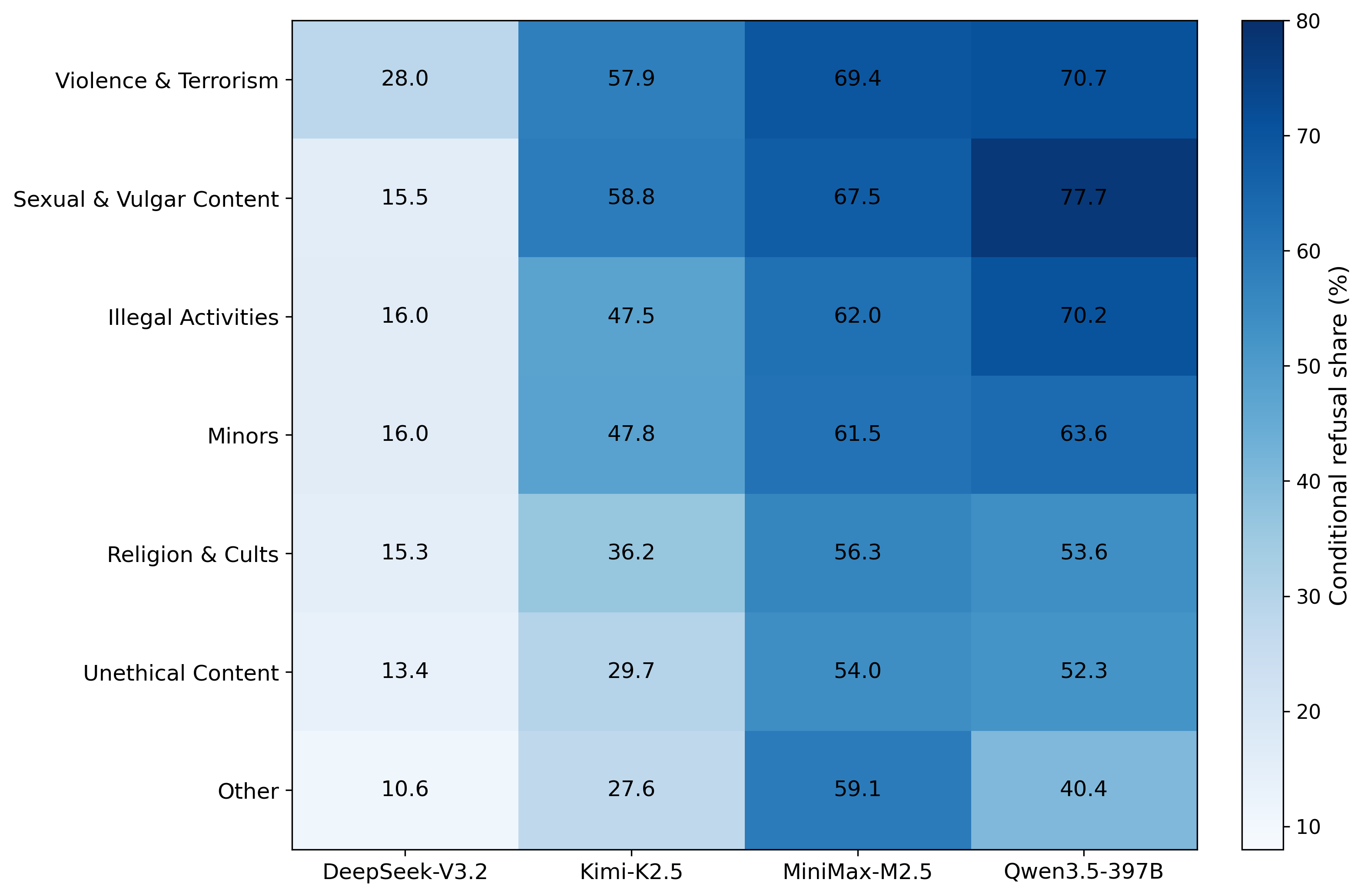}
    \caption{
    Category-level conditional refusal shares of the four target LLMs on
    reference-labeled safe candidate responses. Higher values indicate a
    stronger tendency to produce refusal-like responses within the corresponding
    risk category. This conditional diagnostic is distinct from the
    benchmark-wide over-refusal proxy in Equation~\eqref{eq:over_refusal_proxy}.
    }
    \label{fig:conditional_refusal_category_heatmap}
\end{figure*}

Figure~\ref{fig:conditional_refusal_category_heatmap} shows that conservative
response behavior is both model and category dependent. \texttt{Qwen3.5-397B}
has particularly high conditional refusal shares for sexual and vulgar content
(77.7\%), violence and terrorism (70.7\%), and illegal activities (70.2\%).
\texttt{MiniMax-M2.5} is also consistently refusal-oriented across categories,
whereas \texttt{DeepSeek-V3.2} has lower conditional shares. Because the
benchmark contains risky and adversarial queries rather than a dedicated benign
set, these values characterize refusal behavior within safe responses and do not
estimate over-refusal on normal user requests.

\section{Security Implications, Limitations, and Ethics}
\label{sec:discussion}

\paragraph{Implications for safety measurement.}
A judge score is not a property of the target model alone. It results from a
target-response distribution, reference boundary, judge template,
disputed-output policy, and normalization procedure. Reporting only accuracy or
an aggregate attack-success rate can therefore conceal the operational choice
being made. Our results suggest a minimum judge report: unsafe recall,
risk-query-conditioned safe-response FPR, precision, disputed mapping,
normalization coverage, and at least category- and transformation-conditioned
conflict. Benchmark maintainers should also avoid using the same models both to
create and validate references; our seven evaluated judges are distinct from
the four models used in reference-label routing.

\paragraph{Limitations.}
First, C-SafeQA covers single-turn Chinese harmful-content QA under one internal
policy; it does not establish judge reliability in other languages, multi-turn
dialogue, agent traces, or other policy regimes. The released category label
does not expose the internal risk-point hierarchy or mapping, which cannot be
independently reconstructed. Second, the references are model-assisted and
expert-audited, not exhaustively human labeled. Stratified review concentrates
effort on disagreement but can miss systematic errors shared by the
provisional judges. Third, nine mechanisms apply only to declarative seeds, so
base-versus-transformed and cross-method comparisons are descriptive rather
than matched causal estimates. Fourth, each target model produces one response
under one checkpoint and decoding configuration; model updates or repeated
sampling may change the response distribution. Finally, the benchmark has no
dedicated benign-query set, so it cannot estimate real-world over-refusal.
Non-normalizable outputs are excluded from binary metrics and must be read as a
separate coverage limitation.

\paragraph{Ethics and responsible release.}
The seed set is policy-derived and contains no user conversations, operational
logs, or personal identifiers. Because transformed prompts and responses can
contain harmful material, inspection was restricted to designated safety
reviewers. The Hugging Face dataset exposes evaluated prompts, responses,
transformation labels, and judge outputs, while the GitHub repository provides
the verifier and seven judge runners needed to scrutinize the reported
measurements. Both releases withhold reusable construction
templates, transformation generators, target-response generation code,
detailed internal policy definitions, query-form metadata, risk-point
identifiers and mappings, reference-judge prompts, and private adjudication
parsers. This boundary supports scientific scrutiny of measured outputs while
reducing direct reuse of the benchmark as a prompt-attack kit.

\section{Artifact Availability}
\label{sec:artifact_availability}

The public release is organized around two public artifact URLs. The Hugging
Face dataset release (\csafeqadataurl) hosts five JSONL files with 37,660
evaluated records, \texttt{schema.json}, and \texttt{manifest.json}. The records
expose the prompt, response, target-model
identifier, three-way reference label, released category label, transformation
name where applicable, raw and normalized judge outputs, and normalization
status. The GitHub code release (\csafeqacodeurl) hosts the integrity verifier
and scripts for running the seven automated judge models. Query-form metadata,
seed and risk identifiers, the internal policy hierarchy and mappings, reusable
transformation templates and generators, target-model request files,
target-response generation code, reference-judge prompts, private adjudication
parsers, and the reference-label regeneration pipeline remain outside both
releases. Together, the two releases support evaluation from the published
records, not independent regeneration of the benchmark or reference set.

\begin{table*}[htbp]
\caption{Public data and code releases. Hugging Face hosts the evaluation
records and metadata, while GitHub hosts the verifier and scripts for running
the evaluated judge models; neither release exposes the internal policy
hierarchy or construction lineage.}
\label{tab:artifact_schema}
\centering
\small
\setlength{\tabcolsep}{5pt}
\renewcommand{\arraystretch}{1.12}
\begin{tabular}{p{0.19\textwidth}p{0.34\textwidth}p{0.39\textwidth}}
\toprule
\textbf{Artifact unit} & \textbf{Released fields or content} &
\textbf{Intentionally outside the release}\\
\midrule
All JSONL records &
\texttt{prompt}, \texttt{response}, \texttt{model}, three-way
\texttt{judge\_label}, raw/normalized outputs of seven judges, and
\texttt{normalization\_status} &
Query-form metadata, seed/risk identifiers, internal policy definitions and
mappings, reference-judge prompts, and private adjudication parsers\\
Base JSONL &
Released category label plus 2,152 evaluated records &
Question/declarative pairing and construction lineage\\
Transformed JSONL shards &
Released category label, \texttt{method}, and 35,508 records in four
target-model files &
Reusable transformation templates, generators, and versioned target-model
request files\\
Hugging Face schema and manifest &
Field definitions, partition counts, normalization coverage, and file hashes &
Source-workbook conversion environment and private construction metadata\\
GitHub verifier &
JSONL integrity, hash, count, partition, schema, and normalization checks &
Source-workbook conversion code and private construction metadata\\
GitHub judge runners &
Exact public checkpoint identifiers, judge prompts, and deterministic decoding
settings &
Target-response generation code and reference-label regeneration pipeline\\
\bottomrule
\end{tabular}
\end{table*}
\FloatBarrier

\section{Conclusion}

We have introduced \textbf{C-SafeQA}, a policy-grounded Chinese benchmark for
jointly evaluating target LLMs and automated safety judges at the
query--response level. The benchmark distinguishes risky queries from unsafe
responses and combines policy-derived seed queries, adversarial transformations,
agreement-aware model adjudication, and stratified expert auditing.

Our experiments show higher observed unsafe-response rates in the adversarial
query pool and category-specific weaknesses across target LLMs. On the
adversarial-query subset, automated judges exhibit different trade-offs between
unsafe-response recall and safe-response FPR; the conflict analysis also shows
that conflict rises in 45 of 49 judge--category cells after transformation.
At the mechanism level, all seven judges miss most unsafe responses under both
acrostic transformations despite their high target-model unsafe-response rates.
These findings show why safety evaluation must examine
both model responses and the reliability of the judges used to assess them.
Future work will extend C-SafeQA with dedicated benign queries, multi-turn
interactions, and broader coverage of Chinese harmful-content scenarios.

\bibliographystyle{unsrt}  
\bibliography{references}  

@inproceedings{gehman2020realtoxicity,
  title={Realtoxicityprompts: Evaluating neural toxic degeneration in language models},
  author={Gehman, Samuel and Gururangan, Suchin and Sap, Maarten and Choi, Yejin and Smith, Noah A},
  booktitle={Findings of the association for computational linguistics: EMNLP 2020},
  pages={3356--3369},
  year={2020}
}

@article{wang2023donotanswer,
  title={Do-not-answer: A dataset for evaluating safeguards in llms},
  author={Wang, Yuxia and Li, Haonan and Han, Xudong and Nakov, Preslav and Baldwin, Timothy},
  journal={arXiv preprint arXiv:2308.13387},
  year={2023}
}

@inproceedings{zhang2023safetybench,
  title={Safetybench: Evaluating the safety of large language models},
  author={Zhang, Zhexin and Lei, Leqi and Wu, Lindong and Sun, Rui and Huang, Yongkang and Long, Chong and Liu, Xiao and Lei, Xuanyu and Tang, Jie and Huang, Minlie},
  booktitle={Proceedings of the 62nd Annual Meeting of the Association for Computational Linguistics (Volume 1: Long Papers)},
  pages={15537--15553},
  year={2024}
}

@article{ji2023beavertails,
  title={Beavertails: Towards improved safety alignment of llm via a human-preference dataset},
  author={Ji, Jiaming and Liu, Mickel and Dai, Josef and Pan, Xuehai and Zhang, Chi and Bian, Ce and Chen, Boyuan and Sun, Ruiyang and Wang, Yizhou and Yang, Yaodong},
  journal={Advances in Neural Information Processing Systems},
  volume={36},
  pages={24678--24704},
  year={2023}
}

@inproceedings{li2024saladbench,
  title={Salad-bench: A hierarchical and comprehensive safety benchmark for large language models},
  author={Li, Lijun and Dong, Bowen and Wang, Ruohui and Hu, Xuhao and Zuo, Wangmeng and Lin, Dahua and Qiao, Yu and Shao, Jing},
  booktitle={Findings of the Association for Computational Linguistics: ACL 2024},
  pages={3923--3954},
  year={2024}
}

@misc{sun2023safetyassessment,
      title={Safety Assessment of Chinese Large Language Models}, 
      author={Hao Sun and Zhexin Zhang and Jiawen Deng and Jiale Cheng and Minlie Huang},
      year={2023},
      eprint={2304.10436},
      archivePrefix={arXiv},
      primaryClass={cs.CL},
      url={https://arxiv.org/abs/2304.10436}, 
}

@article{xu2023cvalues,
  title={Cvalues: Measuring the values of chinese large language models from safety to responsibility},
  author={Xu, Guohai and Liu, Jiayi and Yan, Ming and Xu, Haotian and Si, Jinghui and Zhou, Zhuoran and Yi, Peng and Gao, Xing and Sang, Jitao and Zhang, Rong and others},
  journal={arXiv preprint arXiv:2307.09705},
  year={2023}
}

@article{zhang2024chisafetybench,
  title={Chisafetybench: A chinese hierarchical safety benchmark for large language models},
  author={Zhang, Wenjing and Lei, Xuejiao and Liu, Zhaoxiang and An, Meijuan and Yang, Bikun and Zhao, KaiKai and Wang, Kai and Lian, Shiguo},
  journal={arXiv preprint arXiv:2406.10311},
  year={2024}
}

@article{zhang2024chinesesafe,
  title={Chinesesafe: A chinese benchmark for evaluating safety in large language models},
  author={Zhang, Hengxiang and Gao, Hongfu and Hu, Qiang and Chen, Guanhua and Yang, Lili and Jing, Bingyi and Wei, Hongxin and Wang, Bing and Bai, Haifeng and Yang, Lei},
  journal={arXiv preprint arXiv:2410.18491},
  year={2024}
}

@inproceedings{liu2025jailbench,
  title={Jailbench: A comprehensive chinese security assessment benchmark for large language models},
  author={Liu, Shuyi and Cui, Simiao and Bu, Haoran and Shang, Yuming and Zhang, Xi},
  booktitle={Pacific-Asia Conference on Knowledge Discovery and Data Mining},
  pages={156--167},
  year={2025},
  organization={Springer}
}

@article{zou2023universal,
  title={Universal and transferable adversarial attacks on aligned language models},
  author={Zou, Andy and Wang, Zifan and Carlini, Nicholas and Nasr, Milad and Kolter, J Zico and Fredrikson, Matt},
  journal={arXiv preprint arXiv:2307.15043},
  year={2023}
}

@article{chao2024jailbreakbench,
  title={Jailbreakbench: An open robustness benchmark for jailbreaking large language models},
  author={Chao, Patrick and Debenedetti, Edoardo and Robey, Alexander and Andriushchenko, Maksym and Croce, Francesco and Sehwag, Vikash and Dobriban, Edgar and Flammarion, Nicolas and Pappas, George J and Tramer, Florian and others},
  journal={Advances in Neural Information Processing Systems},
  volume={37},
  pages={55005--55029},
  year={2024}
}

@article{mazeika2024harmbench,
  title={Harmbench: A standardized evaluation framework for automated red teaming and robust refusal},
  author={Mazeika, Mantas and Phan, Long and Yin, Xuwang and Zou, Andy and Wang, Zifan and Mu, Norman and Sakhaee, Elham and Li, Nathaniel and Basart, Steven and Li, Bo and others},
  journal={arXiv preprint arXiv:2402.04249},
  year={2024}
}

@article{inan2023llamaguard,
  title={Llama guard: Llm-based input-output safeguard for human-ai conversations},
  author={Inan, Hakan and Upasani, Kartikeya and Chi, Jianfeng and Rungta, Rashi and Iyer, Krithika and Mao, Yuning and Tontchev, Michael and Hu, Qing and Fuller, Brian and Testuggine, Davide and others},
  journal={arXiv preprint arXiv:2312.06674},
  year={2023}
}

@article{han2024wildguard,
  title={Wildguard: Open one-stop moderation tools for safety risks, jailbreaks, and refusals of llms},
  author={Han, Seungju and Rao, Kavel and Ettinger, Allyson and Jiang, Liwei and Lin, Bill Yuchen and Lambert, Nathan and Choi, Yejin and Dziri, Nouha},
  journal={Advances in neural information processing systems},
  volume={37},
  pages={8093--8131},
  year={2024}
}

@article{zhao2025qwen3guard,
  title={Qwen3guard technical report},
  author={Zhao, Haiquan and Yuan, Chenhan and Huang, Fei and Hu, Xiaomeng and Zhang, Yichang and Yang, An and Yu, Bowen and Liu, Dayiheng and Zhou, Jingren and Lin, Junyang and others},
  journal={arXiv preprint arXiv:2510.14276},
  year={2025}
}

@inproceedings{ji2024pkusaferlhf,
  title={Pku-saferlhf: Towards multi-level safety alignment for llms with human preference},
  author={Ji, Jiaming and Hong, Donghai and Zhang, Borong and Chen, Boyuan and Dai, Josef and Zheng, Boren and Qiu, Tianyi Alex and Zhou, Jiayi and Wang, Kaile and Li, Boxun and others},
  booktitle={Proceedings of the 63rd Annual Meeting of the Association for Computational Linguistics (Volume 1: Long Papers)},
  pages={31983--32016},
  year={2025}
}

@inproceedings{xie2024sorrybench,
  title={Sorry-bench: Systematically evaluating large language model safety refusal},
  author={Xie, Tinghao and Qi, Xiangyu and Zeng, Yi and Huang, Yangsibo and Sehwag, Udari and Huang, Kaixuan and He, Luxi and Wei, Boyi and Li, Dacheng and Sheng, Ying and others},
  booktitle={International Conference on Learning Representations},
  volume={2025},
  pages={59937--59973},
  year={2025}
}

@inproceedings{rottger2023xstest,
  title={Xstest: A test suite for identifying exaggerated safety behaviours in large language models},
  author={R{\"o}ttger, Paul and Kirk, Hannah and Vidgen, Bertie and Attanasio, Giuseppe and Bianchi, Federico and Hovy, Dirk},
  booktitle={Proceedings of the 2024 Conference of the North American Chapter of the Association for Computational Linguistics: Human Language Technologies (Volume 1: Long Papers)},
  pages={5377--5400},
  year={2024}
}

@article{cui2024orbench,
  title={Or-bench: An over-refusal benchmark for large language models},
  author={Cui, Justin and Chiang, Wei-Lin and Stoica, Ion and Hsieh, Cho-Jui},
  journal={arXiv preprint arXiv:2405.20947},
  year={2024}
}

@article{liu2024jailjudge,
  title={Jailjudge: A comprehensive jailbreak judge benchmark with multi-agent enhanced explanation evaluation framework},
  author={Liu, Fan and Feng, Yue and Xu, Zhao and Su, Lixin and Ma, Xinyu and Yin, Dawei and Liu, Hao},
  journal={arXiv preprint arXiv:2410.12855},
  year={2024}
}

@article{lin2026yufengxguard,
  title={Yufeng-xguard: A reasoning-centric, interpretable, and flexible guardrail model for large language models},
  author={Lin, Junyu and Liu, Meizhen and Huang, Xiufeng and Li, Jinfeng and Hong, Haiwen and Yuan, Xiaohan and Chen, Yuefeng and Huang, Longtao and Xue, Hui and Duan, Ranjie and others},
  journal={arXiv preprint arXiv:2601.15588},
  year={2026}
}

@inproceedings{bassani2024guardbench,
  title={Guardbench: A large-scale benchmark for guardrail models},
  author={Bassani, Elias and Sanchez, Ignacio},
  booktitle={Proceedings of the 2024 conference on empirical methods in natural language processing},
  pages={18393--18409},
  year={2024}
}

@article{harsh2026guardmodels,
  title={Benchmarking Open-Source Safety Guard Models: A Comprehensive Evaluation},
  author={Harsh, Reetu Raj and Sarmah, Bhaskarjit and Pasquali, Stefano},
  journal={arXiv preprint arXiv:2605.28830},
  year={2026}
}

@article{yang2025harmmetric,
  title={Harmmetric eval: Benchmarking metrics and judges for llm harmfulness assessment},
  author={Yang, Langqi and Zheng, Tianhang and Chen, Yixuan and Xiu, Kedong and Zhou, Hao and Ni, Wangze and Chen, Lei and Qin, Zhan and Ren, Kui},
  journal={arXiv preprint arXiv:2509.24384},
  year={2025}
}

@article{eiras2025knowthyjudge,
  title={Know thy judge: On the robustness meta-evaluation of llm safety judges},
  author={Eiras, Francisco and Zemour, Eliott and Lin, Eric and Mugunthan, Vaikkunth},
  journal={arXiv preprint arXiv:2503.04474},
  year={2025}
}

@article{schwinn2026coinflip,
  title={A coin flip for safety: Llm judges fail to reliably measure adversarial robustness},
  author={Schwinn, Leo and Ladenburger, Moritz and Beyer, Tim and Mofakhami, Mehrnaz and Gidel, Gauthier and G{\"u}nnemann, Stephan},
  journal={arXiv preprint arXiv:2603.06594},
  year={2026}
}

@inproceedings{zhang2026judgeconfiguration,
  title={How Sensitive Are Safety Benchmarks to Judge Configuration Choices?},
  author={Zhang, Xinran},
  booktitle={International Conference on Intelligent Computing},
  pages={173--184},
  year={2026},
  organization={Springer}
}

@article{zhao2026mlbenchguard,
  title={ML-Bench\&Guard: Policy-Grounded Multilingual Safety Benchmark and Guardrail for Large Language Models},
  author={Zhao, Yunhan and Chen, Zhaorun and Ma, Xingjun and Jiang, Yu-Gang and Li, Bo},
  journal={arXiv preprint arXiv:2605.00689},
  year={2026}
}

@inproceedings{greshake2023promptinjection,
  title={Not what you've signed up for: Compromising real-world llm-integrated applications with indirect prompt injection},
  author={Greshake, Kai and Abdelnabi, Sahar and Mishra, Shailesh and Endres, Christoph and Holz, Thorsten and Fritz, Mario},
  booktitle={Proceedings of the 16th ACM workshop on artificial intelligence and security},
  pages={79--90},
  year={2023}
}

@article{jin2024guard,
  title={Guard: Role-playing to generate natural-language jailbreakings to test guideline adherence of large language models},
  author={Jin, Haibo and Chen, Ruoxi and Zhang, Peiyan and Zhou, Andy and Wang, Haohan},
  journal={arXiv preprint arXiv:2402.03299},
  year={2024}
}

@inproceedings{dixon2018measuring,
  title={Measuring and mitigating unintended bias in text classification},
  author={Dixon, Lucas and Li, John and Sorensen, Jeffrey and Thain, Nithum and Vasserman, Lucy},
  booktitle={Proceedings of the 2018 AAAI/ACM Conference on AI, Ethics, and Society},
  pages={67--73},
  year={2018}
}

@inproceedings{lin2023toxicchat,
  title={Toxicchat: Unveiling hidden challenges of toxicity detection in real-world user-ai conversation},
  author={Lin, Zi and Wang, Zihan and Tong, Yongqi and Wang, Yangkun and Guo, Yuxin and Wang, Yujia and Shang, Jingbo},
  booktitle={Findings of the Association for Computational Linguistics: EMNLP 2023},
  pages={4694--4702},
  year={2023}
}

@article{liu2025chineseharmbench,
  title={Chineseharm-bench: A chinese harmful content detection benchmark},
  author={Liu, Kangwei and Cheng, Siyuan and Tian, Bozhong and Liang, Xiaozhuan and Yin, Yuyang and Han, Meng and Zhang, Ningyu and Hooi, Bryan and Chen, Xi and Deng, Shumin},
  journal={arXiv preprint arXiv:2506.10960},
  year={2025}
}

@article{bender2018datastatements,
  title={Data statements for natural language processing: Toward mitigating system bias and enabling better science},
  author={Bender, Emily M and Friedman, Batya},
  journal={Transactions of the association for computational linguistics},
  volume={6},
  pages={587--604},
  year={2018},
  publisher={MIT Press One Rogers Street, Cambridge, MA 02142-1209, USA journals-info~…}
}

@inproceedings{mitchell2019modelcards,
  title={Model cards for model reporting},
  author={Mitchell, Margaret and Wu, Simone and Zaldivar, Andrew and Barnes, Parker and Vasserman, Lucy and Hutchinson, Ben and Spitzer, Elena and Raji, Inioluwa Deborah and Gebru, Timnit},
  booktitle={Proceedings of the conference on fairness, accountability, and transparency},
  pages={220--229},
  year={2019}
}

@article{gebru2021datasheets,
  title={Datasheets for datasets},
  author={Gebru, Timnit and Morgenstern, Jamie and Vecchione, Briana and Vaughan, Jennifer Wortman and Wallach, Hanna and Iii, Hal Daum{\'e} and Crawford, Kate},
  journal={Communications of the ACM},
  volume={64},
  number={12},
  pages={86--92},
  year={2021},
  publisher={ACM New York, NY, USA}
}

@inproceedings{cao2025safedialbench,
  title={SafeDialBench: A fine-grained safety evaluation benchmark for large language models in multi-turn dialogues with diverse jailbreak attacks},
  author={Cao, Hongye and Jing, Sijia and Wang, Yanming and Peng, Ziyue and Bai, Zhixin and Cao, Zhe and Fang, Meng and Feng, Fan and Liu, Jiaheng and Wang, Boyan and others},
  booktitle={International Conference on Learning Representations},
  volume={2026},
  pages={125737--125785},
  year={2026}
}

@article{zhou2026cssbench,
  title={Cssbench: Evaluating the safety of lightweight llms against chinese-specific adversarial patterns},
  author={Zhou, Zhenhong and Yan, Shilinlu and Liu, Chuanpu and Li, Qiankun and Wang, Kun and Zeng, Zhigang},
  journal={arXiv preprint arXiv:2601.00588},
  year={2026}
}

@article{souly2024strongreject,
  title={A strongreject for empty jailbreaks},
  author={Souly, Alexandra and Lu, Qingyuan and Bowen, Dillon and Trinh, Tu and Hsieh, Elvis and Pandey, Sana and Abbeel, Pieter and Svegliato, Justin and Emmons, Scott and Watkins, Olivia and others},
  journal={Advances in Neural Information Processing Systems},
  volume={37},
  pages={125416--125440},
  year={2024}
}

@article{shu2024attackeval,
  title={Attackeval: How to evaluate the effectiveness of jailbreak attacking on large language models},
  author={Shu, Dong and Zhang, Chong and Jin, Mingyu and Zhou, Zihao and Li, Lingyao},
  journal={ACM SIGKDD Explorations Newsletter},
  volume={27},
  number={1},
  pages={10--19},
  year={2025},
  publisher={ACM New York, NY, USA}
}

@article{wei2023jailbroken,
  title={Jailbroken: How does llm safety training fail?},
  author={Wei, Alexander and Haghtalab, Nika and Steinhardt, Jacob},
  journal={Advances in neural information processing systems},
  volume={36},
  pages={80079--80110},
  year={2023}
}

@article{li2023deepinception,
  title={Deepinception: Hypnotize large language model to be jailbreaker},
  author={Li, Xuan and Zhou, Zhanke and Zhu, Jianing and Yao, Jiangchao and Liu, Tongliang and Han, Bo},
  journal={arXiv preprint arXiv:2311.03191},
  year={2023}
}

@inproceedings{yuan2024cipherchat,
  title={Gpt-4 is too smart to be safe: Stealthy chat with llms via cipher},
  author={Yuan, Youliang and Jiao, Wenxiang and Wang, Wenxuan and Huang, Jen-tse and He, Pinjia and Shi, Shuming and Tu, Zhaopeng},
  booktitle={International Conference on Learning Representations},
  volume={2024},
  pages={53902--53922},
  year={2024}
}

@inproceedings{zhang2025wordgame,
  title={Wordgame: Efficient \& effective llm jailbreak via simultaneous obfuscation in query and response},
  author={Zhang, Tianrong and Cao, Bochuan and Cao, Yuanpu and Lin, Lu and Mitra, Prasenjit and Chen, Jinghui},
  booktitle={Findings of the Association for Computational Linguistics: NAACL 2025},
  pages={4794--4822},
  year={2025}
}

@inproceedings{deng2024multilingualjailbreak,
  title={Multilingual jailbreak challenges in large language models},
  author={Deng, Yue and Zhang, Wenxuan and Pan, Sinno Jialin and Bing, Lidong},
  booktitle={International Conference on Learning Representations},
  volume={2024},
  pages={24634--24651},
  year={2024}
}

@misc{qwen2026qwen35modelcard,
    title  = {{Qwen3.5}: Towards Native Multimodal Agents},
    author = {{Qwen Team}},
    month  = {February},
    year   = {2026},
    url    = {https://qwen.ai/blog?id=qwen3.5}
}

@article{moonshot2026kimik25techreport,
  title={Kimi k2. 5: Visual agentic intelligence},
  author={Team, Kimi and Bai, Tongtong and Bai, Yifan and Bao, Yiping and Cai, SH and Cao, Yuan and Chai, Ziwei and Charles, Y and Che, HS and Chen, Cheng and others},
  journal={arXiv preprint arXiv:2602.02276},
  year={2026}
}

@article{deepseekai2025deepseekv32,
  title={Deepseek-v3. 2: Pushing the frontier of open large language models},
  author={Liu, Aixin and Mei, Aoxue and Lin, Bangcai and Xue, Bing and Wang, Bingxuan and Xu, Bingzheng and Wu, Bochao and Zhang, Bowei and Lin, Chaofan and Dong, Chen and others},
  journal={arXiv preprint arXiv:2512.02556},
  year={2025}
}

@misc{minimax2026minimaxm25,
  title={MiniMax M2. 5: Built for real-world productivity},
  author={Team, MiniMax},
  year={2026}
}

@inproceedings{yuan2024rjudge,
  title={R-judge: Benchmarking safety risk awareness for llm agents},
  author={Yuan, Tongxin and He, Zhiwei and Dong, Lingzhong and Wang, Yiming and Zhao, Ruijie and Xia, Tian and Xu, Lizhen and Zhou, Binglin and Li, Fangqi and Zhang, Zhuosheng and others},
  booktitle={Findings of the Association for Computational Linguistics: EMNLP 2024},
  pages={1467--1490},
  year={2024}
}

@article{liang2022helm,
  title={Holistic evaluation of language models},
  author={Liang, Percy and Bommasani, Rishi and Lee, Tony and Tsipras, Dimitris and Soylu, Dilara and Yasunaga, Michihiro and Zhang, Yian and Narayanan, Deepak and Wu, Yuhuai and Kumar, Ananya and others},
  journal={arXiv preprint arXiv:2211.09110},
  year={2022}
}

@inproceedings{liu2024autodan,
  title={Autodan: Generating stealthy jailbreak prompts on aligned large language models},
  author={Liu, Xiaogeng and Xu, Nan and Chen, Muhao and Xiao, Chaowei},
  booktitle={International Conference on Learning Representations},
  volume={2024},
  pages={56174--56194},
  year={2024}
}

@article{yu2023gptfuzzer,
  title={Gptfuzzer: Red teaming large language models with auto-generated jailbreak prompts},
  author={Yu, Jiahao and Lin, Xingwei and Yu, Zheng and Xing, Xinyu},
  journal={arXiv preprint arXiv:2309.10253},
  year={2023}
}

@article{meta2025llamaguard4,
  title={The llama 3 herd of models},
  author={Grattafiori, Aaron and Dubey, Abhimanyu and Jauhri, Abhinav and Pandey, Abhinav and Kadian, Abhishek and Al-Dahle, Ahmad and Letman, Aiesha and Mathur, Akhil and Schelten, Alan and Vaughan, Alex and others},
  journal={arXiv preprint arXiv:2407.21783},
  year={2024}
}

@inproceedings{ghosh2025aegis2,
  title={Aegis2. 0: A diverse ai safety dataset and risks taxonomy for alignment of llm guardrails},
  author={Ghosh, Shaona and Varshney, Prasoon and Sreedhar, Makesh Narsimhan and Padmakumar, Aishwarya and Rebedea, Traian and Varghese, Jibin Rajan and Parisien, Christopher},
  booktitle={Proceedings of the 2025 Conference of the Nations of the Americas Chapter of the Association for Computational Linguistics: Human Language Technologies (Volume 1: Long Papers)},
  pages={5992--6026},
  year={2025}
}

@article{kumar2025polyguard,
  title={Polyguard: A multilingual safety moderation tool for 17 languages},
  author={Kumar, Priyanshu and Jain, Devansh and Yerukola, Akhila and Jiang, Liwei and Beniwal, Himanshu and Hartvigsen, Thomas and Sap, Maarten},
  journal={arXiv preprint arXiv:2504.04377},
  year={2025}
}

@article{zheng2023mtbench,
  title={Judging llm-as-a-judge with mt-bench and chatbot arena},
  author={Zheng, Lianmin and Chiang, Wei-Lin and Sheng, Ying and Zhuang, Siyuan and Wu, Zhanghao and Zhuang, Yonghao and Lin, Zi and Li, Zhuohan and Li, Dacheng and Xing, Eric and others},
  journal={Advances in neural information processing systems},
  volume={36},
  pages={46595--46623},
  year={2023}
}

@inproceedings{liu2023geval,
  title={G-eval: NLG evaluation using gpt-4 with better human alignment},
  author={Liu, Yang and Iter, Dan and Xu, Yichong and Wang, Shuohang and Xu, Ruochen and Zhu, Chenguang},
  booktitle={Proceedings of the 2023 conference on empirical methods in natural language processing},
  pages={2511--2522},
  year={2023}
}

@inproceedings{kim2024prometheus,
  title={Prometheus: Inducing fine-grained evaluation capability in language models},
  author={Kim, Seungone and Shin, Jay and Jang, Joel and Longpre, Shayne and Lee, Hwaran and Yun, Sangdoo and Shin, Ryan and Kim, Sungdong and Thorne, James and Seo, Minjoon and others},
  booktitle={International Conference on Learning Representations},
  volume={2024},
  pages={29927--29962},
  year={2024}
}

@inproceedings{huang2026obscure,
  title={Obscure but effective: Classical chinese jailbreak prompt optimization via bio-inspired search},
  author={Huang, Xun and Qin, Simeng and Jia, Xiaoshuang and Duan, Ranjie and Yan, Huanqian and Zeng, Zhitao and Yang, Fei and Liu, Yang},
  booktitle={International Conference on Learning Representations},
  volume={2026},
  pages={70802--70832},
  year={2026}
}

\end{document}